\documentclass[onecolumn,12pt]{IEEEtran}

\usepackage{epsfig,latexsym}
\usepackage{float}
\usepackage{indentfirst}
\usepackage{amsmath}
\usepackage{amssymb}
\usepackage{times}
\usepackage{subfigure}
\usepackage{psfrag}
\usepackage{cite}
\usepackage{lastpage}
\usepackage{fancyhdr}
\usepackage{color}
 \usepackage{amsthm}
\usepackage{bigints}
\newtheorem{theorem}{Theorem}
\newtheorem{Lemma}{Lemma}
\newtheorem{Corollary}{Corollary}
\newtheorem{lemma}[Lemma]{$\mathbf{Lemma}$}

\newtheorem{corollary}[Corollary]{$\mathbf{Corollary}$}
\begin{document}
\title{ {\huge   The Application of MIMO to   Non-Orthogonal Multiple Access }}

\author{ Zhiguo Ding, \IEEEmembership{Member, IEEE},     Fumiyuki Adachi, \IEEEmembership{Fellow, IEEE},
  and  H. Vincent Poor, \IEEEmembership{Fellow, IEEE}\thanks{
Z. Ding and H. V. Poor  are with the Department of
Electrical Engineering, Princeton University, Princeton, NJ 08544,
USA.   Z. Ding is also with the School of
Computing and Communications, Lancaster
University, LA1 4WA, UK.  F. Adachi is with Graduate School of Eng., Tohoku University,  Sendai, Miyagi, 980-8579 Japan. }\vspace{-0.5em}} \maketitle
\begin{abstract}
 This paper considers the application of multiple-input multiple-output (MIMO) techniques to non-orthogonal multiple access (NOMA) systems. A new design of precoding and detection matrices for MIMO-NOMA is proposed   and its performance is    analyzed for the case with a fixed set of power allocation  coefficients. To further improve the performance gap between MIMO-NOMA and conventional   orthogonal multiple access  schemes,  user pairing is applied to NOMA and its impact on the system performance is characterized. More sophisticated choices of  power allocation coefficients are also proposed to meet various quality of service requirements. Finally computer simulation results are provided to facilitate the performance evaluation of MIMO-NOMA and also demonstrate the accuracy of the developed analytical results.
\end{abstract}\vspace{-1em}
\section{Introduction}
Non-orthogonal multiple access (NOMA) has recently  received considerable  attention as a promising enabling technique in  fifth generation (5G) mobile networks  because of its superior spectral efficiency \cite{6692652} and \cite{6735800}. The key idea of NOMA is to explore the power domain, which has not been used for multiple access (MA) in the previous generations of mobile networks. Specifically NOMA users in one cell are served by a base station (BS) at the same time/code/frequency channel, and their signals are multiplexed by using different power allocation coefficients. The novelty of NOMA comes from the fact that users with poorer channel conditions are allocated   more transmission power.  In this way, these users are able to decode their own messages by treating the others' information as noise, since the power level of their messages is higher. On the other hand, the users with better channel conditions will use the successive interference cancellation (SIC) strategy, i.e., they first decode the messages to the users with poorer channel conditions and then decode their own by removing the other user's information.

The concept of NOMA can be linked to many well-known methods used in previous communication systems. For example, NOMA downlink transmissions   resemble Cover and Thomas's  description   of broadcast channels provided in \cite{Cover1991}. Another example is that the use of SIC has been extensively investigated in conventional multiple-input multiple-output networks, particularly in  V-BLAST systems \cite{738086}. The superimposed messages transmitted in NOMA systems    also resemble the concept of   hierarchical modulation widely used for digital video broadcasting \cite{1433080}. But unlike  these existing techniques, NOMA seeks  to strike a balance between  throughput and fairness. For example, the transmission power allocated to the users in NOMA systems is inversely proportional to their channel conditions, which is important to ensure that all the users are served simultaneously. On the other hand, conventional opportunistic schemes prefer to give more power to users with better channel conditions, which can improve the overall system throughput but deteriorate  fairness.

The impact of path loss on the performance of NOMA has been characterized in \cite{Nomading} by assuming that  users are randomly deployed in a cell, which has demonstrated that NOMA can outperform conventional orthogonal multiple access (OMA) schemes. In \cite{6708131} the implementation of NOMA has been considered in a scenario with two base stations, and the design of uplink NOMA has been proposed in \cite{6933459}. The user fairness  of NOMA has been considered  in \cite{Krikidisnoma} by studying the impact of different choices of power allocation coefficients. In \cite{Zhiguo_pairingnoma} a cognitive radio inspired NOMA scheme was proposed, in which the power allocation coefficients are chosen to meet the predefined users' quality of service (QoS) requirements.

In this paper, we focus on the application of multiple-input multiple-output (MIMO) to NOMA downlink communication systems.  The concept of MIMO-NOMA has been validated by using systematic implementation in \cite{7024800} and \cite{7021086}, which demonstrates that the use of MIMO can outperform   conventional MIMO-OMA. Compared to these existing works, the contributions of this paper are  as follows:

\begin{itemize}
\item We first consider a general NOMA downlink scenario, in which all users participate in NOMA with  a fixed set of power allocation coefficients. A new design of precoding and detection matrices is proposed, and the impact of this design on the performance of NOMA is characterized by using the criteria of outage probabilities and diversity orders. The provided  analytical and numerical results demonstrate that MIMO-NOMA can achieve better outage performance than conventional MIMO-OMA, even for   users that suffer strong co-channel interference.

\item To enlarge the performance gap between MIMO-NOMA and MIMO-OMA, user pairing is applied to NOMA. Analytical results, such as an exact expression for the average sum-rate gap between MIMO-NOMA and MIMO-OMA and its high SNR approximation, are developed. These analytical results demonstrate that the design of user pairing in NOMA systems is very different from conventional user scheduling scenarios. Conventionally it is preferable to schedule the users whose channel conditions are superior, but in the context of NOMA, it is important to schedule users whose channel conditions are very distinct. This is consistent with the findings  obtained   for single-antenna NOMA cases   in \cite{6692652} and \cite{Zhiguo_pairingnoma}.

\item Inspired by the concept of cognitive radio networks, more sophisticated choices for the power allocation coefficients are proposed. Particularly we consider two types of constraints for the power allocation coefficients. One is to meet a predefined QoS requirement, i.e., a user's  rate supported by NOMA is larger than a targeted data rate. The other is to meet a more dynamic QoS requirement, where the user's rate supported by NOMA needs  to be larger than that  supported by conventional OMA. Analytical results are developed for both scenarios to facilitate performance evaluation.
\end{itemize}

\section{System Model With Fixed Power Allocation}\label{section system model}
Consider a downlink  communication scenario, where a BS   equipped with $M$ antennas  communicates with multiple users equipped with $N$ antennas each. To make the NOMA principle applicable to this scenario, the users are randomly grouped into  $M$ clusters with $K$  users in each cluster.
The signals transmitted by the BS are given by
\begin{eqnarray}
\mathbf{x} = \mathbf{P} \tilde{\mathbf{s}},
\end{eqnarray}
where the $M\times 1$ vector $\tilde{\mathbf{s}}$ is given by
\begin{eqnarray}
\tilde{\mathbf{s}} = \begin{bmatrix}\alpha_{1,1}s_{1,1}+\cdots+ \alpha_{1,K}s_{1,K} \\ \vdots \\ \alpha_{M,1}s_{M,1}+\cdots+ \alpha_{M,K}s_{M,K}  \end{bmatrix} \triangleq \begin{bmatrix}
\tilde{s}_1 \\ \vdots \\ \tilde{s}_M
\end{bmatrix}
\end{eqnarray}
where $s_{m,k}$ denotes the information bearing signal to be transmitted to the $k$-th user in the $m$-th cluster, $\alpha_{i,j}$ denotes the NOMA power allocation coefficient,  and the design of the $M\times M$ precoding matrix $\mathbf{P}$ will be discussed in the next section.

Without loss of generality, we focus on the users in the first cluster. The observation at the $k$-th user in the first cluster is given by
\begin{eqnarray}
\mathbf{y}_{1,k} = \mathbf{H}_{1,k}\mathbf{P} \tilde{\mathbf{s}}+\mathbf{n}_{1,k},
\end{eqnarray}
where $\mathbf{H}_{1,k}$ is the $N\times M$ Rayleigh fading  channel matrix from the BS to the $k$-th user in the first cluster, and $\mathbf{n}_{1,k}$ is an  additive Gaussian noise vector. Denote by $\mathbf{v}_{1,k}$   the detection vector used this user. After applying this detection vector, the signal model can be rewritten as follows:
\begin{eqnarray}
\mathbf{v}_{1,k}^H\mathbf{y}_{1,k} &=& \mathbf{v}_{1,k}^H\mathbf{H}_{1,k}\mathbf{P} \tilde{\mathbf{s}}+\mathbf{v}_{1,k}^H\mathbf{n}_{1,k}.
\end{eqnarray}
Denote the $i$-th column of $\mathbf{P}$ by $\mathbf{p}_i$. The above signal model can be rewritten as follows:
\begin{align}
\mathbf{v}_{1,k}^H\mathbf{y}_{1,k} &= \mathbf{v}_{1,k}^H\mathbf{H}_{1,k}\mathbf{p}_1 \left(\alpha_{1,1}s_{1,1}+\cdots+ \alpha_{1,K}s_{1,K}\right)\\ \nonumber &+\sum_{m=2}^{M}\mathbf{v}_{1,k}^H\mathbf{H}_{1,k}\mathbf{p}_{m} \tilde{ {s}}_{m}+\mathbf{v}_{1,k}^H\mathbf{n}_{1,k}.
\end{align}
The channel conditions are crucial to the implementation of NOMA.  Without loss of generality, we assume that the effective channel gains are ordered  as follows:
\begin{align}\label{order}
 |\mathbf{v}_{1,1}^H\mathbf{H}_{1,1}\mathbf{p}_1|^2 \geq \cdots \geq |\mathbf{v}_{1,K}^H\mathbf{H}_{1,K}\mathbf{p}_1|^2,
\end{align}
and following the principle of NOMA, the users' power allocation coefficients are ordered as follows:
\[
\alpha_{1,1}\leq \cdots \leq \alpha_{1,K}.
\]
In this section, constant   power allocation coefficients will be considered, and more sophisticated choices will be used in Section \ref{section CRNOMA}.  It is worth pointing out that optimizing   power allocation according to instantaneous channel conditions can be used to further improve the performance of MIMO-NOMA, which is beyond  the scope of this paper.

Based on the above signal model, the signal-to-interference-plus-noise (SINR) for the $K$-th ordered user in the first cluster is given by{\small
\begin{align}
&SINR_{1,K} =\\ \nonumber & \frac{|\mathbf{v}_{1,K}^H\mathbf{H}_{1,K}\mathbf{p}_1 |^2 \alpha_{1,K}^2 }{ \sum^{K-1}_{l=1}|\mathbf{v}_{1,K}^H\mathbf{H}_{1,K}\mathbf{p}_1 |^2 \alpha_{1,l}^2 +\sum_{m=2}^{M}|\mathbf{v}_{1,K}^H\mathbf{H}_{1,K}\mathbf{p}_{m}|^2 +|\mathbf{v}_{1,K}|^2\frac{1}{\rho}},
\end{align}}
$\hspace{-1em}$ where $\rho$ denotes the transmit signal to noise ratio (SNR).

 The $k$-th  user, $1<k<K$, needs to decode the messages to the users with poorer channel conditions first, before detecting   its own. The messages $s_{1,j}$, $K\geq j\geq (k+1)$, will be detected  at the $k$-th user with the following SINR:{\small
\begin{align}
&SINR^j_{1,k} =\\ \nonumber & \frac{|\mathbf{v}_{1,k}^H\mathbf{H}_{1,k}\mathbf{p}_1 |^2 \alpha_{1,j}^2 }{ \sum^{j-1}_{l=1}|\mathbf{v}_{1,k}^H\mathbf{H}_{1,k}\mathbf{p}_1 |^2 \alpha_{1,l}^2 +\sum_{m=2}^{M}|\mathbf{v}_{1,k}^H\mathbf{H}_{1,k}\mathbf{p}_{m}|^2 +|\mathbf{v}_{1,k}|^2\frac{1}{\rho}}.
\end{align}}
$\hspace{-1em}$ If the message $s_{1,j}$ can be decoded successfully, i.e., $\log(1+SINR^j_{1,k})>R_{1,j}$,
then it will be removed from the $k$-th user's observation, where $R_{i,j}$ denotes the $j$-th user's targeted data rate.     This SIC will be carried out until the $k$-th user's own message is decoded with the SINR, $SINR^k_{1,k}$.

The first user in the first cluster   needs to decode all the other users' messages with $SINR^j_{1,1}$, $K\geq j \geq 2$. If successful, it will decode its own message with the following SINR:
\begin{align}
SINR^1_{1,1} =  \frac{|\mathbf{v}_{1,1}^H\mathbf{H}_{1,1}\mathbf{p}_1 |^2 \alpha_{1,1}^2 }{  \sum_{m=2}^{M}|\mathbf{v}_{1,1}^H\mathbf{H}_{1,1}\mathbf{p}_{m}|^2 +|\mathbf{v}_{1,1}|^2\frac{1}{\rho}}.
\end{align}
The design of the precoding and detection  matrices will be discussed in the following section.

\section{Design of Precoding and Detection Matrices}\label{section performance analysis}
To completely remove  inter-cluster interference, the   precoding and detection matrices need to satisfy the following constraints:
\begin{align}\label{vp}
\mathbf{v}_{i,k}^H\mathbf{H}_{i,k}\mathbf{p}_{m}=0,
\end{align}
for any $m\neq i$.

In order to reduce system overhead caused by acquiring  channel state information (CSI) at the BS, it is assumed that the BS does not have the global CSI\footnote{It is worth pointing out that the BS still needs to know the order of the users' effective channel gains in order to implement NOMA as shown in \eqref{order}, but this imposes a much less demanding requirement  compared to  knowing all the users' channel matrices at the BS. }, which leads to the following choice of  $\mathbf{P}$:
\[
\mathbf{P} = \mathbf{I}_M,
\]
where $\mathbf{I}_M$ is the $M\times M $ identity matrix. The above choice means  that the BS broadcasts the users' messages without manipulating them.
The advantage of this choice is that it avoids asking the users to feedback all their CSI to the BS, which  consumes significant system overhead.

With this choice of $\mathbf{P}$, the constraints on the detection matrices in \eqref{vp} become
\begin{align}
\mathbf{v}_{i,k}^H\mathbf{h}_{m,ik}=0,
\end{align}
where $\mathbf{h}_{m,ik}$ is the $m$-th column of $\mathbf{H}_{i,k}$. Therefore at the $k$-th user in the $i$-th cluser, the constraints can be rewritten as follows:
\begin{align}\nonumber
\mathbf{v}_{i,k}^H\underset{\tilde{\mathbf{H}}_{i,k}}{\underbrace{\begin{bmatrix} \mathbf{h}_{1,ik}&\cdots &\mathbf{h}_{i-1,ik}&\mathbf{h}_{i+1,ik}&\cdots&\mathbf{h}_{M,ik}\end{bmatrix}}}=0.
\end{align}
Note that the dimension of $\tilde{\mathbf{H}}_{i,k}$ is $N\times (M-1)$ since it is a submatrix of $\mathbf{H}_{i,k}$ formed by removing one column.
As a result, $\mathbf{v}_{i,k}$ can be obtained from the null space of $\tilde{\mathbf{H}}_{i,k}$, i.e.,
\begin{align}
\mathbf{v}_{i,k} = \mathbf{U}_{i,k}\mathbf{z}_{i,k},
\end{align}
where $\mathbf{U}_{i,k}$ contains all the left singular vectors of $\tilde{\mathbf{H}}_{i,k}$ corresponding to zero singular values, and $\mathbf{z}_{i,k}$ is a  $(N-M+1)\times 1$ normalized vector to be optimized later. In order to ensure the existence of $\mathbf{v}_{i,k}$, $N\geq M$ is assumed.

By using the above precoding and detection matrices, the SINR for  the $K$-th user in the first cluster is given by
\begin{align}
&SINR_{1,K} = \frac{|\mathbf{v}_{1,K}^H\mathbf{h}_{1,1K}|^2 \alpha_{1,K}^2 }{ \sum^{K-1}_{l=1}|\mathbf{v}_{1,K}^H\mathbf{h}_{1,1K} |^2 \alpha_{1,l}^2  +|\mathbf{v}_{1,K}|^2\frac{1}{\rho}},
\end{align}
where inter-cluster interference has been removed.

At the $k$-th users, $1<k<K$, the messages $s_{1,j}$, $K\geq j\geq (k+1)$, will be detected   with the following SINR:
\begin{align}
&SINR^j_{1,k} = \frac{|\mathbf{v}_{1,k}^H\mathbf{h}_{1,1k}  |^2 \alpha_{1,j}^2 }{ \sum^{j-1}_{l=1}|\mathbf{v}_{1,k}^H\mathbf{h}_{1,1k} |^2 \alpha_{1,l}^2  +|\mathbf{v}_{1,k}|^2\frac{1}{\rho}}.
\end{align}
If successful,  $s_{j,1}$ will be removed from the $k$-th user's observation, and SIC will be carried out until its own message is decoded with the SINR, $SINR^k_{1,k}$.

The first user in the first cluster will  decode the other users' messages with $SINR^j_{1,1}$, $K\geq j \geq 2$. If successful, it will decode its own message with the following SINR:
\begin{align}
SINR^1_{1,1} =  \rho \frac{|\mathbf{v}_{1,1}^H\mathbf{h}_{1,11}  |^2 \alpha_{1,1}^2 }{   |\mathbf{v}_{1,1}|^2 }.
\end{align}
As can be observed from the above SINR expressions, $\mathbf{z}_{i,k}$ determines the SINRs through $|\mathbf{v}_{i,k}^H\mathbf{h}_{i,ik}  |^2$. Therefore, one possible   choice of $\mathbf{z}_{i,k}$ can be obtained by using  maximal  radio combining (MRC) approach. Particularly, the choice of $\mathbf{z}_{i,k}$ based on MRC is given by
\begin{eqnarray}
\mathbf{z}_{i,k} = \frac{\mathbf{U}_{i,k}^H \mathbf{h}_{i,ik}}{|\mathbf{U}_{i,k}^H \mathbf{h}_{i,ik}|}.
\end{eqnarray}
The following theorem provides an exact expression for the outage probability achieved by MIMO-NOMA and its high SNR approximation.
\begin{theorem}\label{theorem1}
Assume that the users in each cluster are ordered as in \eqref{order}. With MIMO-NOMA, the outage probability experienced by the $k$-th ordered  user in the $i$-th cluster is given by
\begin{align}
\mathrm{P}^o_{i,k}  =  \sum^{k-1}_{p=0}{k-1 \choose p}   \frac{(-1)^p K!\left[\frac{\gamma\left(N-M+1,\epsilon_{i,k}^*
\right)}{(N-M)!} \right]^{K-k+p+1} }{(K-k)!(k-1)!(K-k+p+1) },
\end{align}
 if $\alpha_{i,j}^2 >\beta_{i,j}$, for all $k\leq j\leq K$,  otherwise $\mathrm{P}^o_{i,K}=1$, where $\epsilon_{i,k}=2^{R_{i,k}}-1$, $\beta_{i,k}= \epsilon_{i,k}\sum^{K-1}_{k=1}  \alpha_{i,k}^2$, $\gamma(\cdot)$ denotes the incomplete gamma function,   $\epsilon^*_{i,k} = \max \left\{\frac{\epsilon_{i,K}}{\rho\left(\alpha_{i,K}^2 -\beta_{i,K}\right) }, \cdots, \frac{\epsilon_{i,k}}{\rho\left(\alpha_{i,k}^2 -\beta_{i,k}\right) } \right\}$, for $2\leq k \leq K$
 and
$ \epsilon^*_{i,1} = \max \left\{\frac{\epsilon_{i,K}}{\rho\left(\alpha_{i,K}^2 -\beta_{i,K}\right) }, \cdots, \frac{\epsilon_{i,2}}{\rho\left(\alpha_{i,2}^2 -\beta_{i,2}\right) },  \frac{\epsilon_{i,1}}{\rho\alpha_{i,1}^2  }\right\}$.
A high SNR approximation for the outage probability is given by
\begin{align}
\mathrm{P}^o_{i,k} \approx  \frac{ K!\left[
\frac{ \left(\epsilon_{i,k}^*\right)^{N-M+1}}{(N-M+1)!  }
  \right]^{K-k+1} }{(K-k)!(k-1)!(K-k+1)  }.
\end{align}
\end{theorem}
\begin{proof}
Please refer to the appendix.
\end{proof}
A benchmarking scheme based on conventional MIMO-OMA can be described as follows. The MIMO-OMA transmission consists of $K$ time slot. During each time slot, $M$ users, one from each cluster, are served simultaneously  based on the same manner as described for MIMO-NOMA. As a result, the SINR at the $k$-th user in the $i$-th cluster is given by
\begin{align}
&SINR_{i,k} = \frac{|\mathbf{v}_{i,k}^H\mathbf{H}_{i,k}\mathbf{p}_i |^2 }{ \sum_{m=1, m\neq i}^{M}|\mathbf{v}_{i,k}^H\mathbf{H}_{i,k}\mathbf{p}_{m}|^2 +|\mathbf{v}_{i,k}|^2\frac{1}{\rho}}.
\end{align}
Note that   the   MRC detection vector used for MIMO-NOMA is also applicable to MIMO-OMA. In addition consider that  the users in one cluster are also sorted as in \eqref{order}.  The outage probability achieved by this  version of MIMO-OMA can be obtained in the following corollary straightforwardly by following the steps in the proof for Theorem \ref{theorem1}.
\begin{corollary} \label{corollary}
Assume that the users in each cluster are ordered as in \eqref{order}. By applying conventional MIMO-OMA, the outage probability experienced by  the $k$-th ordered  user in the $i$-th cluster is given by
\begin{align}
\mathrm{P}^o_{i,k}  =  \sum^{k-1}_{p=0}{k-1 \choose p}   \frac{(-1)^p K!\left[\frac{\gamma\left(N-M+1,\phi_{i,k}
\right)}{(N-M)!} \right]^{K-k+p+1} }{(K-k)!(k-1)!(K-k+p+1) },
\end{align}
where $\phi_{i,k}=\frac{2^{KR_{i,k}}-1}{\rho}$.
A high SNR approximation for the outage probability is given by
\begin{align}
\mathrm{P}^o_{1,k} \approx  \frac{ K!\left[
\frac{ \left(\phi_{i,k}\right)^{N-M+1}}{(N-M+1)!  }
  \right]^{K-k+1} }{(K-k)!(k-1)!(K-k+1)  }.
\end{align}
\end{corollary}
As can be observed from Theorem \ref{theorem1}  and Corollary \ref{corollary}, MIMO-NOMA can achieve a diversity gain of $(N-M+1)(K-k+1))$, the same as conventional MIMO-OMA. But this diversity gain is achieved by allowing   all the $K$ users from the same cluster to share the same bandwidth resource, which yields better spectral efficiency. For example, the simulation results provided in Section \ref{section numerical result} demonstrate that  MIMO-NOMA can achieve a smaller outage probability compared to conventional NOMA. The superior spectral efficiency of MIMO-NOMA can also be demonstrated by the fact that  it can realize a larger sum rate,  as shown in  the following section when the impact of user pairing is investigated.

\section{The Impact of User Pairing}\label{section user pairing}
User pairing has the potential to reduce the complexity of NOMA systems. Specifically the   users in one cluster  can be divided into groups with fewer users in each group. A hybrid multiple access scheme can be used, where NOMA will be implemented among the users within each  group, and conventional OMA can be used for inter-group multiple access. In addition to reducing system complexity, user pairing/grouping can also significantly  increase the performance gain of NOMA over conventional MIMO-OMA,  as shown in the following.

In order to obtain some insightful analytical results, we focus on the case in which two users are paired together for performing NOMA in each cluster. Particularly  the $n$-th and $k$-th  ordered users from each  cluster are scheduled to perform  NOMA, where the $n$-th user has a better channel condition, i.e., $n<k$. By using the same choices of the precoding and detection matrices, the SNR for the $k$-th user in the first cluster is given by
\begin{align}
&SNR_{1,k} =  \frac{|\mathbf{v}_{1,k}^H\mathbf{h}_{1,1k}|^2 \alpha_{1,k}^2 }{  |\mathbf{v}_{1,k}^H\mathbf{h}_{1,1k}|^2 \alpha_{1,n}^2  + |\mathbf{v}_{1,k}|^2 \frac{1}{\rho}},
\end{align}
and the SNR at  the $n$-th user is given by
\begin{align}
&SNR_{1,n} = \rho \frac{|\mathbf{v}_{1,n}^H\mathbf{h}_{1,1n}|^2 \alpha_{1,n}^2 }{  |\mathbf{v}_{1,n}|^2 },
\end{align}
conditioned on the event that  the $n$-th user can decode the other user's information correctly.
Note that the power allocation coefficients satisfy $\alpha_{1,n}^2+\alpha_{1,k}^2=1$.

We are particularly interested in the sum-rate gap between MIMO-NOMA and conventional MIMO, which is given by
\begin{align}
\Delta & \triangleq  \sum^{M}_{i=1}\left[\log \left(1+SNR_{i,k} \right) +\log \left(1+SNR_{i,n} \right)\right]   \\ \nonumber &- \frac{1}{2}\sum^{M}_{i=1} \left[ \log \left(1+\rho|\mathbf{v}_{i,k}^H\mathbf{h}_{i,ik}|^2  \right) +\log \left(1+|\mathbf{v}_{i,n}^H\mathbf{h}_{i,in}|^2  \right)\right].
\end{align}

Following the same definitions used in the proof for Theorem \ref{theorem1}, the average sum rate gap can be expressed as follows:
\begin{align}\nonumber
&\mathcal{E}\left\{\Delta\right\}  =  M\mathcal{E}\left\{ \log \left(1+SNR_{1,k} \right) +\log \left(1+SNR_{1,n} \right) \right\}   \\ \nonumber &- \frac{M}{2}\mathcal{E}\left\{ \log \left(1+\rho|\mathbf{v}_{1,k}^H\mathbf{h}_{1,1k}|^2  \right) +\log \left(1+|\mathbf{v}_{1,n}^H\mathbf{h}_{1,1n}|^2  \right) \right\}\\ \nonumber&=
M\mathcal{E}\left\{ \log \left(1+  \frac{x_{k} \alpha_{1,k}^2 }{  x_k \alpha_{1,n}^2  + \frac{1}{\rho}} \right) +\log \left(1+x_n\alpha_{1,n}^2 \rho\right) \right\}   \\ \nonumber &- \frac{M}{2}\mathcal{E}\left\{ \log \left(1+\rho x_k \right) +\log \left(1+\rho x_n \right) \right\},
\end{align}
where $x_k=|\mathbf{v}_{1,k}^H\mathbf{h}_{1,1k}|^2$ for notational simplicity.

After some manipulations, we can write 
\begin{align}\nonumber
&\mathcal{E}\left\{\Delta\right\} =\frac{M}{2}
\mathcal{E}\left\{ \log \left(  1+\rho x_{k}    \right)\right\} +M\mathcal{E}\left\{\log \left(1+\rho x_n\alpha_{1,n}^2 \right) \right\}   \\   &- M \mathcal{E}\left\{\log\left(1+ \rho x_k \alpha_{1,n}^2\right)\right\}   -\frac{M}{2}\mathcal{E}\left\{\log \left(1+\rho x_n \right) \right\}.
\end{align}

The key for evaluating the rate gap $\mathcal{E}\left\{\Delta\right\}$ is to characterize $\mathcal{E}\left\{\log \left(1+x_n\phi \right) \right\} $ which can be calculated as follows:
\begin{align}
&\mathcal{E}\left\{\log \left(1+x_n\phi \right) \right\}
\\ \nonumber =&- \int^{\infty}_{0}\log \left(1+x\phi \right)  d (1-F_{x_n}(x))
\\ \nonumber =&\frac{\phi}{\ln 2}\int^{\infty}_{0}  \frac{1-F_{x_n}(x)}{1+x\phi}dx.
\end{align}

By applying the cumulative distribution function (CDF) of the channel gain,  $x_n$, provided in \eqref{pdfd xxk} in the proof for Theorem~\ref{theorem1}, the sum rate gap can be expressed as follows: {\small
\begin{align}
&\mathcal{E}\left\{\log \left(1+x_n\phi \right) \right\}
\\ \nonumber =&\frac{\phi}{\ln 2}\int^{\infty}_{0}\hspace{-0.5em}  \frac{1-\gamma_n \int^{x}_{0}  f_{\tilde{x}_k}(x)[F_{\tilde{x}_k}(x) ]^{K-n} [1-F_{\tilde{x}_k}(x) ]^{n-1} dy}{1+x\phi}dx
\\ \nonumber =&\frac{\phi}{\ln 2}\int^{\infty}_{0}  \frac{1}{1+x\phi}\left(1-\sum^{n-1}_{p=0}{n-1 \choose p}\gamma_n\right.\\ \nonumber &\left. \times (-1)^p \frac{[F_{\tilde{x}_k}(x) ]^{K-n+p+1}}{K-n+p+1}\right)   dx,
\end{align}}
$\hspace{-1em}$ where $\gamma_n = \frac{K!}{(K-n)!(n-1)!}$ and the CDF $F_{\tilde{x}_k}(x)$ is obtained following the density function in  \eqref{pdf of xk}. By using the above equation and with some straightforward manipulations, the ergodic rate gap can be obtained   in the following lemma.
\begin{lemma}\label{lemma1}
Suppose  that the $n$-th and $k$-th users are grouped to perform MIMO-NOMA. The average sum rate gap between MIMO-NOMA and conventional MIMO-OMA is given by
\begin{align}\nonumber
\mathcal{E}\left\{\Delta\right\} &= \frac{M}{2}\varphi(k,\rho) +M\varphi(n,\rho  \alpha_{1,n}^2)   - M \varphi(k,\rho  \alpha_{1,n}^2) \\ \label{lemma eq} & -\frac{M}{2}\varphi(n,\rho  ) ,
\end{align}
where
\begin{align}
\varphi(n,\phi) =&\frac{\phi}{\ln 2}\int^{\infty}_{0}  \frac{1}{1+x\phi}\left(1-\sum^{n-1}_{p=0}{n-1 \choose p}\gamma_n\right.\\ \nonumber &\left. \times (-1)^p \frac{\left[\frac{\gamma(N-M+1,x)}{(N-M)!} \right]^{K-n+p+1}}{K-n+p+1}\right)   dx.
\end{align}
\end{lemma}

While the analytical result in Lemma \ref{lemma1} can be used to replace Monte-Carlo simulations for performance evaluation, this is  still quite complicated due to the  integrals and   special functions.  In the following, some case studies will be carried out in order to obtain some insight into  MIMO-NOMA.

\subsection*{Case studies  for the sum-rate gain of MIMO-NOMA}
In this subsection, we focus on two extreme cases as described in the following:
\begin{itemize}
\item Case I:  In each cluster, pair the   user having the worst channel condition with the one having the  best channel condition, i.e., $n=1$ and $k=K$.
\item Case II: In each cluster, pair the   user having the best channel condition with the one having the second best channel condition, i.e., $n=1$ and $k=2$.
\end{itemize}
In conventional MA systems, scheduling users with better channel conditions is beneficial for improving system throughput, but we can show that NOMA has a behavior different from conventional MA.
\begin{lemma}\label{lemma 2}
For the case with $N=M=2$, $n=1$ and $k=K$, the average sum-rate gap between MIMO-NOMA and MIMO-OMA is given by \begin{align}\nonumber
\mathcal{E}\left\{\Delta\right\}  & = -  \log(e) e^{\frac{K}{\rho}}\mathbf{E_i}\left(-\frac{K}{\rho}\right) \\ \nonumber &+\frac{2}{ \ln 2}   \left(     \sum^{K}_{l=1}{K \choose l}(-1)^l e^{\frac{l}{\rho  \alpha_{1,1}^2}}  \mathbf{E_i}\left(-\frac{l}{\rho  \alpha_{1,1}^2}\right)  \right)   \\ \nonumber &+2  \log(e) e^{\frac{K}{\rho\alpha_{1,1}^2}}\mathbf{E_i}\left(-\frac{K}{\rho\alpha_{1,1}^2}\right)   \\ \label{lemma1 eq1} &-\frac{1}{ \ln 2}   \left(     \sum^{K}_{l=1}{K \choose l}(-1)^l e^{\frac{l}{\rho}}  \mathbf{E_i}\left(-\frac{l}{\rho}\right)  \right) ,
\end{align}
where $\mathbf{E_i}(\cdot)$ denotes the exponential integral function. At high SNR, the gap can be approximated as follows:
 \begin{align}
\mathcal{E}\left\{\Delta\right\} \approx  \log   K  +          \sum^{K}_{l=1}{K \choose l}(-1)^l    \log l
   .
\end{align}
For the case with $N=M=2$, $n=1$ and $k=2$, the average sum-rate gap between MIMO-NOMA and MIMO-OMA is given by \eqref{figure eq1}.
\begin{figure*}
\begin{align}\label{figure eq1}
\mathcal{E}\left\{\Delta\right\}   & =\frac{1}{ \ln 2}   \left(K  \sum^{K-1}_{p=1} {K-1 \choose p}(-1)^p   e^{\frac{p}{\rho}}  \mathbf{E_i}\left(-\frac{p}{\rho}\right)  - (K-1)    \sum^{K}_{l=1}{K \choose l}(-1)^l e^{\frac{l}{\rho}}  \mathbf{E_i}\left(-\frac{l}{\rho}\right)  \right)   \\ \nonumber &+\frac{2}{ \ln 2}   \left(     \sum^{K}_{l=1}{K \choose l}(-1)^l e^{\frac{l}{\rho  \alpha_{1,1}^2}}  \mathbf{E_i}\left(-\frac{l}{\rho  \alpha_{1,1}^2}\right)  \right)   \\ \nonumber &-  \frac{2}{ \ln 2}   \left(K  \sum^{K-1}_{p=1} {K-1 \choose p}(-1)^p   e^{\frac{p}{\rho  \alpha_{1,1}^2}}  \mathbf{E_i}\left(-\frac{p}{\rho  \alpha_{1,1}^2}\right)  -(K-1)    \sum^{K}_{l=1}{K \choose l}(-1)^l e^{\frac{l}{\rho  \alpha_{1,1}^2}}  \mathbf{E_i}\left(-\frac{l}{\rho  \alpha_{1,1}^2}\right)  \right)    \\ \nonumber &-\frac{1}{ \ln 2}   \left(     \sum^{K}_{l=1}{K \choose l}(-1)^l e^{\frac{l}{\rho}}  \mathbf{E_i}\left(-\frac{l}{\rho}\right)  \right) .
\end{align}
\end{figure*}
At high SNR, the average gap can be approximated as follows:
\begin{align}\label{hign snr part 2}
\mathcal{E}\left\{\Delta\right\}  &\approx K   \left(- \sum^{K-1}_{p=1} {K-1 \choose p}(-1)^p      \log  p \right. \\ \nonumber &\left. +   \sum^{K}_{l=1}{K \choose l}(-1)^l \log l  \right).
\end{align}
\end{lemma}
\begin{proof}
Please refer to the appendix.
\end{proof}
Define $\varpi(k)=\sum^{k}_{l=1}{k \choose l}(-1)^l    \log l$ which is a mono-increasing function of $k$. Lemma \ref{lemma 2} shows that,  at high SNR, the sum-rate gap for Case I can be approximated   as $ (\log   K  +         \varpi(K))$, which means that the larger $K$ is, the more gain MIMO-NOMA can offer compared to conventional MIMO-OMA.
On the other hand, numerical results show that  the value of $k(\varpi(k)-\varpi(k-1))$ quickly goes to zero by increasing $k$, which means the sum-rate gain offered by MIMO-NOMA for Case II is diminishing with increasing $K$.  These two extreme cases   demonstrate that careful user pairing is critical for MIMO-NOMA to outperform conventional MIMO-OMA. Detailed numerical analysis will be provided in Section \ref{section numerical result}.

\section{Cognitive Radio Inspired MIMO-NOMA}\label{section CRNOMA}
In the previous sections, fixed choices of power allocation coefficients have been considered, and in this section,  more sophisticated choices will be used. Without loss of generality, we focus on the same case as in Section \ref{section user pairing}, i.e., the $n$-th and $k$-th users from each cluster are selected to perform NOMA and the $k$-th user has poorer channel conditions, i.e., $n<k$.

An important observation is that there is a dilemma in NOMA systems for choosing $\alpha_{1,k}$. From the perspective of the overall system throughput, an ideal choice of $\alpha_{1,k}$ is $\alpha_{1,k}=0$, i.e., all power is allocated to the user with better channel conditions. But this choice completely ignores the user fairness, and in this section we focus on two   choices of $\alpha_{i,k}$ inspired by the concept of cognitive radio networks.

 \subsection{To meet a fixed QoS requirement }
 Consider that there is a targeted SINR threshold to ensure the QoS requirement at the $k$-th user, i.e., $SINR_{i,k}\geq \epsilon_{i,k}$. This SINR requirement imposes  the following constraint on the  power coefficient $\alpha_{i,k}^2$:
\begin{align}\label{cr constraint}
 1
\geq \alpha_{i,k}^2 \geq  \frac{ \epsilon_{i,k}\left( |\mathbf{v}_{i,k}^H\mathbf{h}_{i,ik}|^2 +\frac{1}{\rho}\right)}{|\mathbf{v}_{i,k}^H\mathbf{h}_{i,ik}|^2 (1+\epsilon_{i,k})}.
\end{align}
In this paper, we will simply set $\alpha_{i,k}$ as follows:
\begin{align}\label{choice1}
\alpha_{i,k}^2 =  \min\left\{1,\frac{ \epsilon_{i,k}\left( |\mathbf{v}_{i,k}^H\mathbf{h}_{i,ik}|^2 +\frac{1}{\rho}\right)}{|\mathbf{v}_{i,k}^H\mathbf{h}_{i,ik}|^2 (1+\epsilon_{i,k})}\right\}.
\end{align}
 This choice of $\alpha_{i,k}^2 $ means that the BS will give the $k$-th user the minimal transmission power needed  to meet this user's QoS requirement, and then allocate the remaining power to the $n$-th user.

 The outage probability experienced at the $k$-th user is equal  to  $\mathrm{P}(\alpha_{i,k}=1)$ or equivalently  $\mathrm{P}\left(\frac{ \epsilon_{i,k}\left( |\mathbf{v}_{i,k}^H\mathbf{h}_{i,ik}|^2 +\frac{1}{\rho}\right)}{|\mathbf{v}_{i,k}^H\mathbf{h}_{i,ik}|^2 (1+\epsilon_{i,k})}>1\right)$ , i.e., the $k$-th user's targeted data rate cannot be supported even if the BS allocates all the power to this user.   Following the proof of Theorem~\ref{theorem1}, it is straightforward to show that a  diversity order of $(N-k+1)(N-M+1)$ is achievable at the $k$-th user, because \[
 \mathrm{P}\left(\frac{ \epsilon_{i,k}\left( |\mathbf{v}_{i,k}^H\mathbf{h}_{i,ik}|^2 +\frac{1}{\rho}\right)}{|\mathbf{v}_{i,k}^H\mathbf{h}_{i,ik}|^2 (1+\epsilon_{i,k})}>1\right) = \mathrm{P}\left(|\mathbf{v}_{i,k}^H\mathbf{h}_{i,ik}|^2<\frac{\epsilon_{i,k}}{\rho}\right).
 \]
 The following theorem demonstrates the achievable diversity order at the $n$-th user.
 \begin{lemma}\label{lemma 3}
 With the cognitive radio inspired power allocation coefficient $\alpha_{i,k}$ in \eqref{choice1}, a diversity gain of $(N-M+1)(K-k+1)$ is achievable at the $n$-th user.
 \end{lemma}
\begin{proof}Please refer to the appendix.
\end{proof}
It is important to point out that the diversity gain at the $n$-th user is constrained by the $k$-th user's channel condition due to the use of  \eqref{choice1}, which is consistent with the finding in \cite{Zhiguo_pairingnoma}. Recall that  cognitive radio inspired NOMA  with signal-antenna nodes can  achieve a diversity  of $(K-k+1)$ for both users \cite{Zhiguo_pairingnoma}. Therefore one advantage of  MIMO-NOMA is that a larger diversity order can be achieved. In addition, the use of MIMO-OFDM can ensure that   more users are served  simultaneously.
%

\subsection{To meet a dynamic QoS constraint}
Another choice for the QoS requirement at the $k$-th user is to ensure the following constraint:
\begin{align}\label{constraint xx}
\log\left(1+\frac{|\mathbf{v}_{1,k}^H\mathbf{h}_{1,1k}|^2  \alpha_{1,k}^2 }{ |\mathbf{v}_{1,k}^H\mathbf{h}_{1,1k} |^2 \alpha_{1,n}^2 + \frac{1}{\rho}}\right)>\frac{1}{2}\log\left(1+ |\mathbf{v}_{1,k}^H\mathbf{h}_{1,1k}|^2   \rho\right),
\end{align}
which means that the $k$-th user is willing to perform NOMA with the $n$-th user only if it can achieve a larger rate compared to the case with    conventional MIMO-OMA.

With the same notation as before, the above constraint can be expressed as follows:
\begin{align}\label{constaint}
  \frac{(1+x_k\rho)^2 }{ (1+x_k\alpha_{1,n}^2\rho)^2} > \left(1+ x_k  \rho\right),
\end{align}
from which the constrain for the power coefficient $\alpha_{1,n}^2$ can be obtained as follows:
\begin{align}
0\leq \alpha_{1,n} \leq \sqrt{\frac{\sqrt{1+\rho x_k} -1}{\rho x_k}}.
\end{align}
Note that  $0\leq \frac{\sqrt{1+\rho x_k} -1}{\rho x_k}\leq 1$ for arbitrary choices of $x_k$, and therefore $\frac{\sqrt{1+\rho x_k} -1}{\rho x_k}$ is a feasible choice. So we will set
\begin{align}
\label{coeff}
\alpha_{1,n} = \sqrt{\frac{\sqrt{1+\rho x_k} -1}{\rho x_k}},
 \end{align}which is     the maximal value of the power allocation coefficient given the constraint in  \eqref{constaint}.

 We first focus on the impact of this power coefficient on the outage probability at the $k$-th user, which   is given by
\begin{align}\label{kuser}
\mathrm{P}^k_o &= \mathrm{P}\left(\log\left(1+\frac{x_k \alpha_{1,k}^2 }{ x_k \alpha_{1,n}^2 + \frac{1}{\rho}}\right) <R_{1,k} \right)\\ \nonumber &=  \mathrm{P}\left(\log \sqrt{1+\rho x_k}<R_{1,k} \right).
\end{align}
An important conclusion from \eqref{kuser} is  that the use of the   power coefficient in \eqref{coeff} ensures that the $k$-th user experiences   exactly the same outage probability as the case with conventional MIMO-OMA.  This observation is expected since the choice of $\alpha_{1,n}$ is to ensure the constraint in \eqref{constraint xx}, i.e., the $k$-th user's rate should  not be reduced because of the use of NOMA. Following     steps similar to those used  in the previous section, it is straightforward to show that the diversity gain of this user is $(N-M+1)(K-k+1)$.

Because the  expression for the power allocation coefficient in \eqref{coeff} is very complicated, an exact expression for the outage probability achieved at the $n$-th user is difficult  to find, but the {\it achievable} diversity gain can still be obtained as shown in the following lemma.
\begin{lemma}\label{theorem2}
In the proposed CR-MIMO-NOMA system with the dynamic QoS constraint in \eqref{constraint xx}, a diversity order of $(N-M+1)(K-k+1)$ is achievable by the $n$-th user.
\end{lemma}
\begin{proof}Please refer to the appendix.
\end{proof}

It is worth pointing out that the diversity order provided in Lemma \ref{theorem2} is only an achievable one. After carrying out computer simulations, we observe that this diversity lower bound is tight for the case of $R_{1,k}>1$, and a diversity order larger than $(N-M+1)(K-k+1)$ can be achieved for $0\leq R_{1,k}\leq 1$.  A possible reason for  this is that a loose  bound is used to get the achievable diversity gain for the case of  $0\leq R_{1,k}\leq 1$, as shown  in \eqref{eqx1}.

\section{Numerical Results}\label{section numerical result}
In this section computer simulation results will be used to demonstrate the performance of MIMO-NOMA and also verify the accuracy of the developed analytical results. For notational simplicity, we omit the index of the cluser, e.g., $R_{1,k}$ is denoted by $R_{k}$.

\begin{figure}[!htp]
\begin{center} \includegraphics[width=0.47\textwidth]{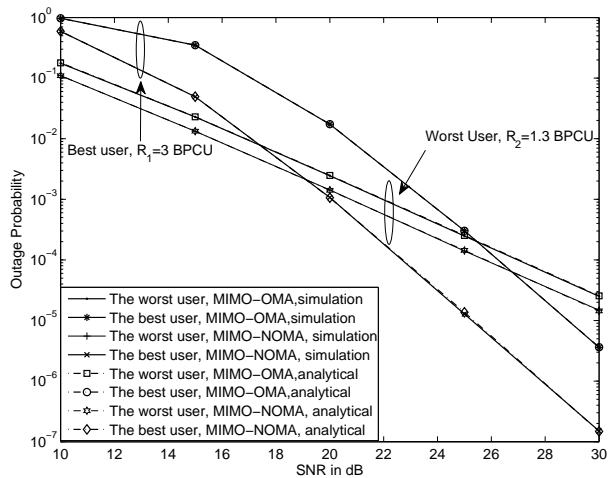}
\end{center}
\vspace*{-3mm} \caption{MIMO-NOMA with a fixed set of power coefficients.   $M=2$, $N=3$ and $K=2$. $\alpha_1^2=\frac{1}{4}$ and $\alpha_2^2=\frac{3}{4}$. BPCU denotes bit per channel use.   }\label{fig fix1}
\end{figure}

\begin{figure}[!htp]
\begin{center} \includegraphics[width=0.47\textwidth]{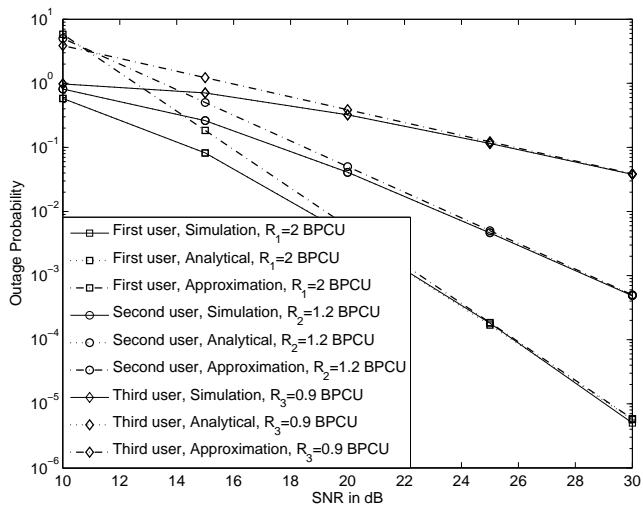}
\end{center}
\vspace*{-3mm} \caption{MIMO-NOMA with a fixed set of power coefficients.   $M=3$, $N=3$ and $K=3$. $\alpha_1^2=\frac{1}{6}$, $\alpha_2^2=\frac{1}{3}$ and $\alpha_3^2=\frac{1}{2}$. }\label{fig fix2 1}
\end{figure}

In Figs. \ref{fig fix1} and \ref{fig fix2 1} the performance of MIMO-NOMA with fixed power allocation coefficients is studied first. Particularly,  Fig. \ref{fig fix1} considers the case in which there are four users grouped into two clusters, with two users in each cluster.  Fig. \ref{fig fix2 1}  considers the case in which there are three clusters, with three users in each cluster. All users in each cluster will participate in NOMA. Fig. \ref{fig fix1} confirms the accuracy of the analytical results developed in Theorem \ref{theorem1} and Corollary \ref{corollary}. In addition this figure also demonstrates that MIMO-NOMA can achieve better outage  performance than MIMO-OMA though both realize the same diversity gain. Fig. \ref{fig fix2 1}   demonstrates the accuracy of the high SNR approximation results developed in Theorem \ref{theorem1}. In particular, one observation from this figure is that different users experience different diversity orders, which confirms the diversity order results developed in Theorem \ref{theorem1}.

\begin{figure}[!htp]
\begin{center} \subfigure[ Case I: $n=1$ and $k=K$]{\includegraphics[width=0.47\textwidth]{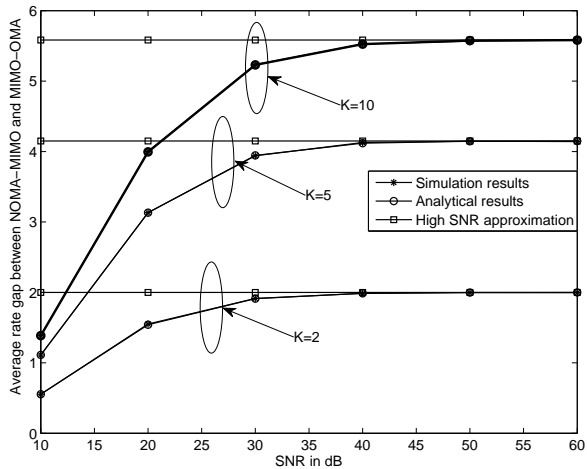}}
\subfigure[Case II: $n=1$ and $k=2$]{
\includegraphics[width=0.47\textwidth  ]{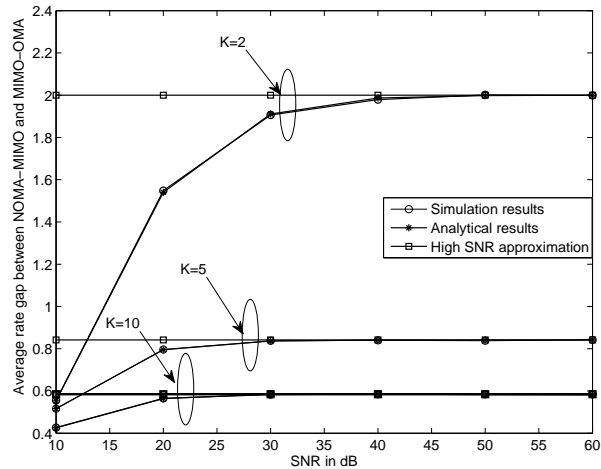}}
\end{center}
\vspace*{-3mm} \caption{The performance gap offered by MIMO-NOMA,  $M=2$   $N=2$ and $\alpha_1^2=\frac{1}{4}$.   }\label{fig ergodic 1}
\end{figure}

In Fig. \ref{fig ergodic 1} the impact of user pairing is demonstrated by using the sum-rate gap between MIMO-NOMA and MIMO-OMA. As can be seen from both sub-figures, the exact expression for the average sum-rate gap developed in Lemma \ref{lemma1} matches the simulation results perfectly, and the approximation result developed in the lemma provides a tight bound at  high SNR. Comparing Fig. \ref{fig ergodic 1}(a) to Fig. \ref{fig ergodic 1}(b), one can observe that the impact of $K$ on the performance gap is much different. In Fig. \ref{fig ergodic 1}(a),  increasing the number of the users in each group, $K$, can significantly improve the performance gap between MIMO-NOMA and MIMO-OMA. Specifically a gain of $2$ bits per channel use (BPCU) can be obtained when there are $2$ users in each group, and this gap can double  when there are $5$ users in each group. The reason for this performance gain is  because we schedule the best user and the worst user, i.e., $n=1$ and $k=K$, and the two selected users' channel information becomes very different when increasing $K$,  which is beneficial to the implementation of NOMA. On the other hand,  Fig. \ref{fig ergodic 1}(b) demonstrates that the performance gain of MIMO-NOMA is diminishing with increasing $K$. This is because the user with the best channel conditions and the one with the second best channel conditions are scheduled. When increasing $K$,  the two users' channel conditions become significantly similar, which will reduce the performance gain of NOMA.

\begin{figure}[!htp]
\begin{center} \includegraphics[width=0.47\textwidth]{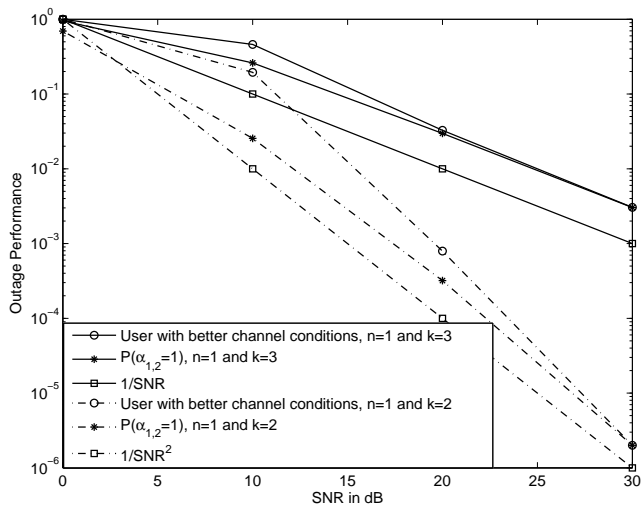}
\end{center}
\vspace*{-3mm} \caption{Cognitive radio inspired MIMO-NOMA with a fixed power constraint, $\epsilon_{1,k}=1$, $M=2$, $N=2$, $K=3$ and $R_1=2$ BPCU.  }\label{fig gap 1}
\end{figure}

In Fig. \ref{fig gap 1}, the performance of cognitive radio (CR) inspired MIMO-NOMA for meeting  the fixed QoS requirement in \eqref{cr constraint} is studied. In this figure three types of curves are provided, one for $\mathrm{P}_n^o$ as studied in Lemma \ref{lemma 3}, one for $\mathrm{P}(\alpha_{2}=1)$, and one for $\frac{1}{\rho^{(N-M+1)(K-k+1)}}$. The last is provided to demonstrate the achievable diversity order.  As can be seen in the figure, the curves for $\mathrm{P}_n^o$  are parallel to the ones for $\frac{1}{\rho^{(N-M+1)(K-k+1)}}$, which demonstrates that the achievable diversity order obtained in Lemma \ref{lemma 3} is tight. An interesting observation from the figure is that   $\mathrm{P}(\alpha_{2}=1)$ is a tight  lower bound of $\mathrm{P}_n^o$, particularly at high SNR. This is because CR-MIMO-NOMA tends to satisfy the $k$-th user's QoS first and therefore the event  $\alpha_{2}^2=1$, i.e., the BS allocates all the power to the $k$-th user,  is dominant among the three types of events described in the proof for Lemma \ref{lemma 3}.

\begin{figure}[!htp]
\begin{center} \subfigure[ $K=3$, $n=1$ and $R_{n}=R_{k}=2$ BPCU]{\includegraphics[width=0.47\textwidth]{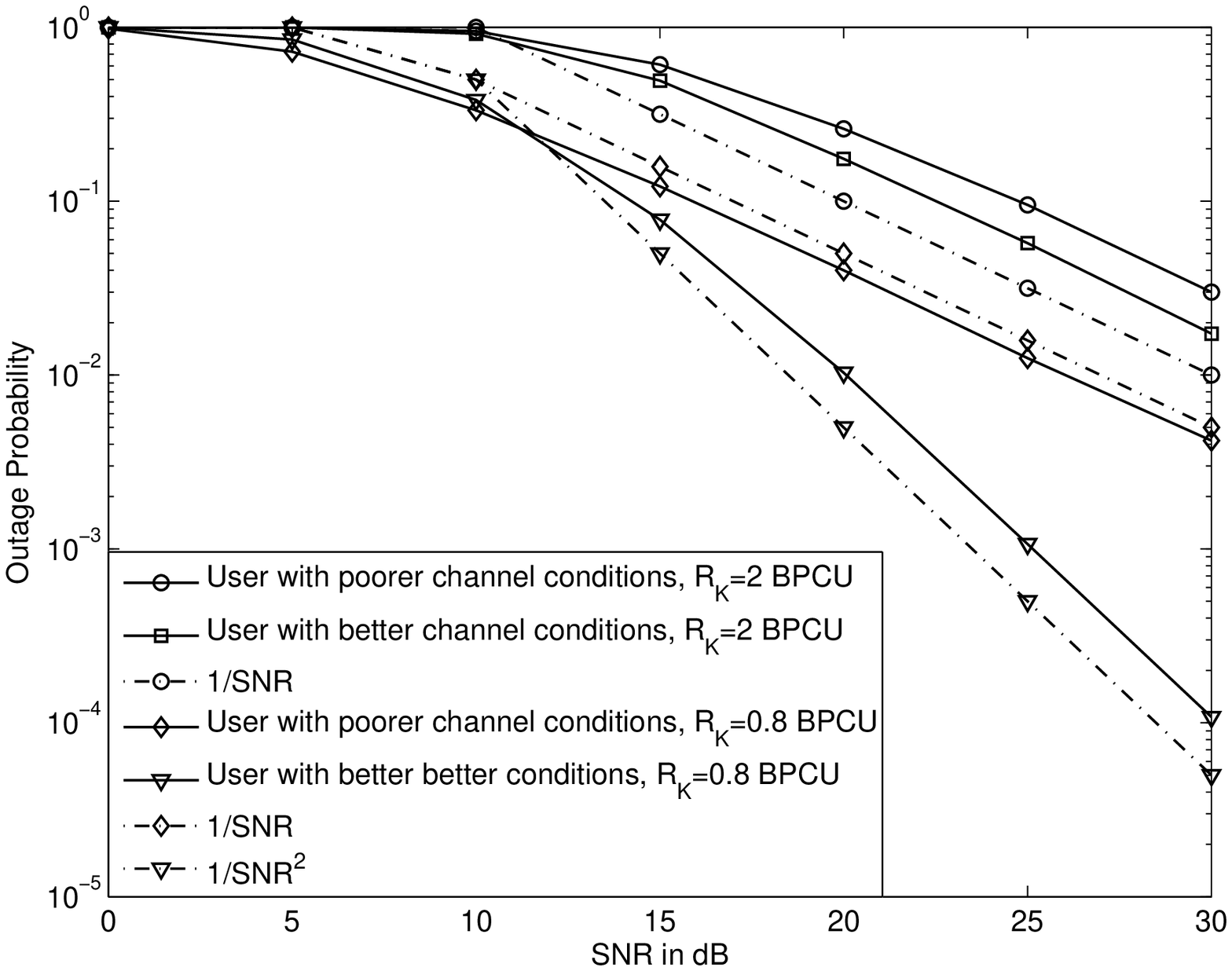}}
\subfigure[$K=2$, $n=1$, $k=K$ and $R_{n}$ =2 BPCU]{
\includegraphics[width=0.47\textwidth  ]{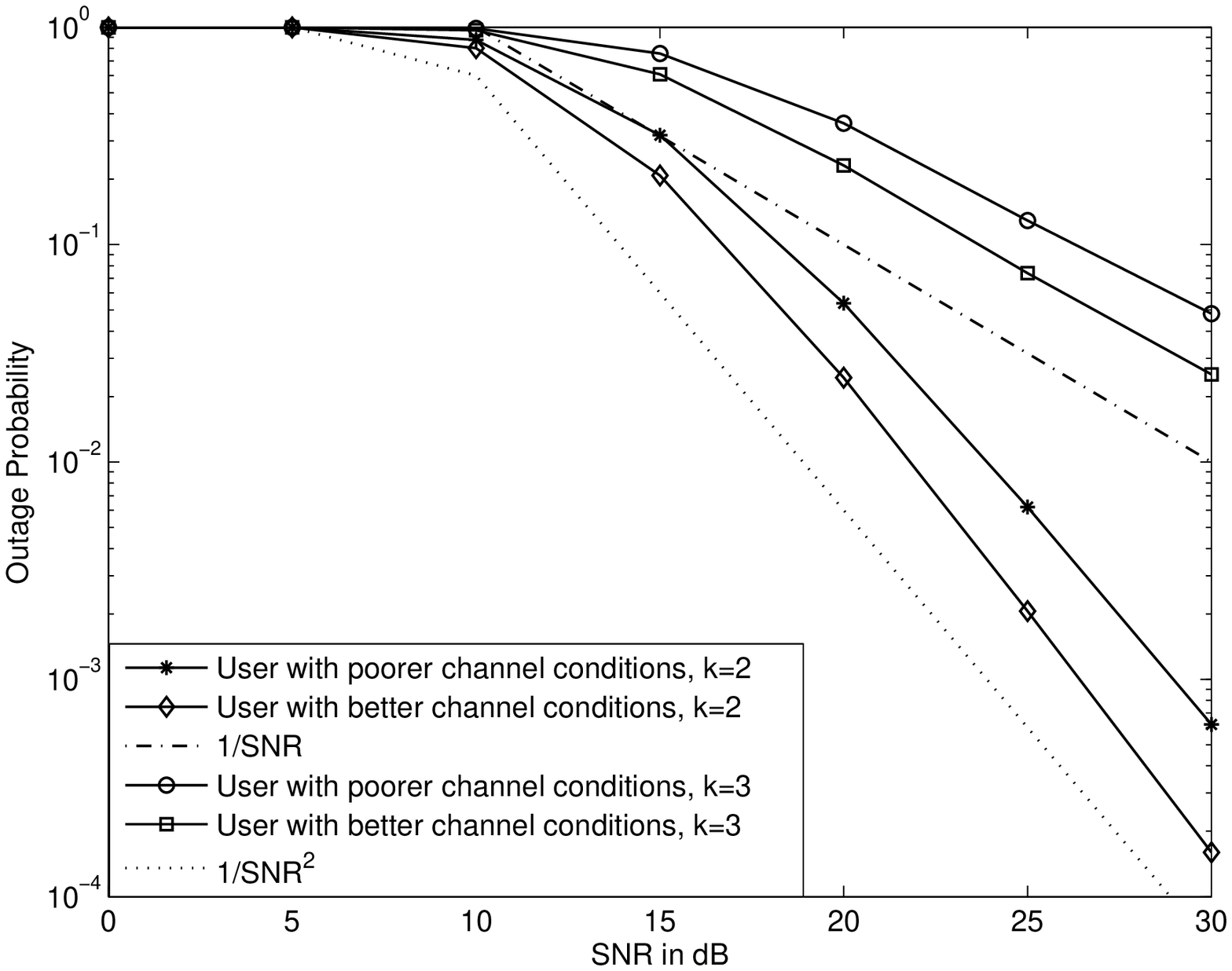}}
\end{center}
\vspace*{-3mm} \caption{CR with a dynamic power constraint,  $M=2$, $N=2$  }\label{fig crfix 1}
\end{figure}

Finally, the performance of CR-MIMO-NOMA in meeting  the dynamic  QoS requirement   in \eqref{constraint xx} is investigated in Fig. \ref{fig crfix 1}. Again the  curves  for $\frac{1}{\rho^{2}}$ and $\frac{1}{\rho^{2}}$ are provided to facilitate the analysis of diversity orders.  Both   sub-figures demonstrate that a diversity order of $(N-M+1)(K-k+1)$ is achievable regardless of the choice of $R_k$, which confirms the accuracy of Lemma \ref{theorem2}. Furthermore, this   diversity order of $(N-M+1)(K-k+1)$ can be tight depending on the choice of $R_k$. For example, in Fig. \ref{fig crfix 1}(a), when $R_k=2$ BPCU, the curves for the outage probability for the user with better channel conditions  are always parallel to the ones for $\frac{1}{\rho^{(N-M+1)(K-k+1)}}$. In general, our carried out simulation studies reveal that  the   diversity order of $(N-M+1)(K-k+1)$ is exactly what CR-MIMO-NOMA can realize in the case of $R_k>1$. However, in the case of $0\leq R_k \leq 1$, a diversity gain larger than $(N-M+1)(K-k+1)$ can be achieved, as shown in Fig. \ref{fig crfix 1}(b). As discussed in Section \ref{section CRNOMA}, the reason for this is because the upper bound used in the proof for Lemma \ref{theorem2} is loose in the case of $0\leq R_k \leq 1$.

\section{Conclusion}
In this paper, we have studied  the application of MIMO to NOMA systems. A new design of precoding and detection matrices for MIMO-NOMA has been proposed,   and its performance has been analyzed.   To further improve the performance gap between MIMO-NOMA and conventional   OMA, the use of  user pairing has been considered in  NOMA systems and its impact on the system performance has also been characterized. The cognitive radio  inspired choices for  power allocation coefficients have also been  proposed to meet various QoS requirements.  Simulation results have been provided to demonstrate the accuracy of the developed analytical results. In this paper, it is assumed that users have  been randomly divided into multiple groups, and an important future direction is   to study the design of low complexity approaches for dynamic clustering/grouping in MIMO-NOMA systems.

\appendices
\section{Proof for Theorem \ref{theorem1}}
The proof can be completed in four steps. \vspace{-1em}
\subsection{Density function of effective channel gains}Without loss of generality, we only focus on the users in the first cluster.
First recall that these users have been ordered according to the   criterion in \eqref{order} which can be rewritten as follows:
\begin{align}\label{order1}
 x_1 \geq \cdots \geq x_K,
\end{align}
where $x_k\triangleq  |\mathbf{v}_{1,k}^H\mathbf{h}_{1,1k}|^2$. Define $\tilde{x}_k$ as the unordered counterpart of $x_k$. Given the choice of $\mathbf{v}_{1,k} = \mathbf{U}_{1,k}\mathbf{z}_{1,k}$ and $\mathbf{z}_{1,k} = \frac{\mathbf{U}_{1,k}^H \mathbf{h}_{1,1k}}{|\mathbf{U}_{1,k}^H \mathbf{h}_{1,1k}|}$, we have
\[
|\mathbf{v}_{1,k}^H\mathbf{h}_{1,1k}|^2 = \left(\frac{|\mathbf{U}_{1,k}^H \mathbf{h}_{1,1k}|^2}{|\mathbf{U}_{1,k}^H \mathbf{h}_{1,1k}|}\right)^2 = |\mathbf{U}_{1,k}^H \mathbf{h}_{1,1k}|^2.
\]

An important observation is that  $\mathbf{U}_{1,k}$ contains the $(N-M+1)$ orthogonal singular vectors, i.e.,
\[
\mathbf{U}_{1,k}^H\mathbf{U}_{1,k}=\mathbf{I}_{N-M+1},
\] and also note  that $\mathbf{U}_{1,k}$ is independent of  $\mathbf{h}_{1,1k}$. Therefore $\mathbf{v}_{1,k}^H\mathbf{h}_{1,1k}$ represents a unitary transformation of a complex Gaussian vector, which means that $\mathbf{v}_{1,k}^H\mathbf{h}_{1,1k}$  is still an $(N-M+1)\times 1$ complex Gaussian vector \cite{Raomaxitrx}. Therefore this unordered variable, $\tilde{x}_k$, follows the chi-square distribution, and thus the probability density function (pdf) of $\tilde{x}_k$ is given by
\begin{eqnarray}\label{pdf of xk}
f_{\tilde{x}_k}(x) = \frac{e^{-x}}{(N-M)!}x^{N-M},
\end{eqnarray}
and its CDF  is $F_{\tilde{x}}(x) = \int^{x}_{0}f_{\tilde{x}_k}(y)dy$.
Therefore the ordered variable, $x_k$, in \eqref{order1} follows the following pdf \cite{David03}:
\begin{eqnarray}\label{pdfd xxk}
f_{{x}_k}(x) = \frac{K!f_{\tilde{x}_k}(x)[F_{\tilde{x}_k}(x) ]^{K-k} [1-F_{\tilde{x}_k}(x) ]^{k-1}}{(K-k)!(k-1)!}.
\end{eqnarray}

\vspace{-1em}
\subsection{A unified outage probability expression}
Because the users in one cluster carry out different detection  strategies, the outage probabilities achieved by different users will be evaluated separately first and then a unified expression for these probabilities will be developed.
\subsubsection{Outage probability at the user with the worst channel condition}
The outage probability  for  the $K$-th user in the first cluster is given by
\begin{align}
&\mathrm{P}\left(SINR_{1,K}<\epsilon_{1,K}\right) \\ \nonumber &= \mathrm{P}\left(\frac{x_K \alpha_{1,K}^2 }{ x_K\sum^{K-1}_{k=1}  \alpha_{1,k}^2  +\frac{1}{\rho}}<\epsilon_{1,K}\right).
\end{align}
The above outage probability can be written as follows:
\begin{align}\label{outage expressions1}
&\mathrm{P}\left(SINR_{1,K}<\epsilon_{1,K}\right) \\ \nonumber &= {\small \left\{\begin{array}{ll} \mathrm{P}\left( x_K   <\frac{\epsilon_{1,K}}{\rho\left(\alpha_{1,K}^2 -\beta_{1,K}\right) } \right),& \text{if}\quad  \alpha_{1,K}^2 >\beta_{1,K}\\ 1, &\text{otherwise}\end{array}\right..}
\end{align}
where $\beta_{1,K}= \epsilon_{1,K}\sum^{K-1}_{k=1}  \alpha_{1,k}^2$.

\subsubsection{Outage probability at the $k$-th user, $1<k<K$}
The $k$-th user needs to decode the $j$-th user's message, $j>k$, before detecting  its own message.
The overall outage probability for the $k$-th user to decode its own message can be expressed as follows:
\begin{align}
\mathrm{P}^o_{1,k} &= 1- \mathrm{P}\left(SINR^j_{1,k}>\epsilon_{1,j}, \forall j \in \{k, \cdots, K\}\right)\\ \nonumber
&=1- \mathrm{P}\left(\frac{x_k \alpha_{1,j}^2 }{ x_k\sum^{j-1}_{l=1}  \alpha_{1,l}^2  +\frac{1}{\rho}}>\epsilon_{1,j}, \forall j \in \{k, \cdots, K\}\right).
\end{align}

 Following  steps similar to those used in the previous subsection, the above probability can be rewritten as follows:
\begin{align}&\mathrm{P}\left(\frac{x_k \alpha_{1,j}^2 }{ x_k\sum^{j-1}_{l=1}  \alpha_{1,l}^2  +\frac{1}{\rho}}>\epsilon_{1,j}, \forall j \in \{k, \cdots, K\}\right)\\ \nonumber & = \left\{\begin{array}{ll} \mathrm{P}\left( x_k   >\frac{\epsilon_{1,j}}{\rho\left(\alpha_{1,j}^2 -\beta_{1,j}\right) }, \forall j \in \{k, \cdots, K\} \right),& \text{if {\bf C1}}   \\ 0, &\hspace{-1em}\text{otherwise}\end{array}\right.\hspace{-0.5em}.
\end{align}
where the condition, {\bf C1}, denotes   $\alpha_{1,j}^2 > \beta_{1,j}$, for all $k\leq j\leq K$, and $\beta_{1,j}= \epsilon_{1,j}\sum^{j-1}_{l=1}  \alpha_{1,l}^2$.

Define $\epsilon^*_{1,k} = \max \left\{\frac{\epsilon_{1,j}}{\rho\left(\alpha_{1,j}^2 -\beta_{1,j}\right) }, k\leq j \leq K\right\}$. The outage probability can   be expressed as follows:
\begin{align}\label{outage expressions}
\mathrm{P}^o_{1,k} &= \left\{\begin{array}{ll}  \mathrm{P}\left( x_k <\epsilon_{1,k}^*\right),& \text{if {\bf C1}}   \\ 1, &\text{otherwise}\end{array}\right..
\end{align}
It is interesting to observe that the expression in   \eqref{outage expressions1} is a special case of  \eqref{outage expressions}. It is straightforward to evaluate that the outage probability expressions in \eqref{outage expressions} can also be  used for the user with the best channel condition by letting $\epsilon^*_{1,1} = \max \left\{\frac{\epsilon_{1,K}}{\rho\left(\alpha_{1,K}^2 -\beta_{1,K}\right) }, \cdots, \frac{\epsilon_{1,2}}{\rho\left(\alpha_{1,2}^2 -\beta_{1,2}\right) },  \frac{\epsilon_{1,1}}{\rho\alpha_{1,1}^2  }\right\}$.

\subsection{Obtaining an exact expression for the outage probability }
When the conditions, $\alpha_{1,k}^2 \geq \beta_{1,k}$, are satisfied,    the outage probability is given by
\begin{align}&
\mathrm{P}^o_{1,k}  =  \sum^{k-1}_{p=0}{k-1 \choose p} (-1)^p \frac{K!}{(K-k)!(k-1)!}\\ \nonumber &\times \int^{\epsilon_{1,k}^* }_{0}  f_{\tilde{x}_k}(x)\left[F_{\tilde{x}_k}\left(x\right) \right]^{K-k+p} dx \\ \nonumber &=  \sum^{k-1}_{p=0}{k-1 \choose p} \frac{(-1)^p K!\left[F_{\tilde{x}_k}\left(\epsilon_{1,k}^*
\right) \right]^{K-k+p+1} }{(K-k)!(k-1)!(K-k+p+1)} .
\end{align}

By applying the CDF of the unsorted variable $\tilde{x}_k$, we obtain
\begin{align}\nonumber
\mathrm{P}^o_{1,k}  = \sum^{k-1}_{p=0}{k-1 \choose p} \frac{(-1)^p K!\left[\int^{\epsilon_{1,k}^*}_{0}f_{\tilde{x}_k}(y)dy   \right]^{K-k+p+1} }{(K-k)!(k-1)!(K-k+p+1)} .
\end{align}
By applying the incomplete gamma function, the exact expression of the outage probability can be obtained as in the theorem.

\subsection{High SNR approximations}
By applying the series expansion of the incomplete gamma function \cite{GRADSHTEYN}, the outage probability can be first expanded as in \eqref{eq fie}.
\begin{figure*}
\begin{eqnarray}\label{eq fie}
\mathrm{P}^o_{1,k}  & =&  \sum^{k-1}_{p=0}{k-1 \choose p} (-1)^p K! \frac{\left[(N-M)!\left(1-e^{-\epsilon_{1,k}^*} \sum^{N-M}_{q=0}
\frac{\left(\epsilon_{1,k}^*\right)^q}{q!  }
\right) \right]^{K-k+p+1} }{(K-k)!(k-1)!(K-k+p+1) ((N-M)!)^{K-k+p+1}}
\\\nonumber
& =&  \sum^{k-1}_{p=0}{k-1 \choose p}  (-1)^p K!\frac{\left[(N-M)!\left(1-e^{-\epsilon_{1,k}^*} \left(e^{\epsilon_{1,k}^*} -\sum_{q=N-M+1}^{\infty}
\frac{\left(\epsilon_{1,k}^*\right)^q}{q! }\right)
\right) \right]^{K-k+p+1} }{(K-k)!(k-1)!(K-k+p+1) ((N-M)!))^{K-k+p+1}}.
\end{eqnarray}
\end{figure*}
At high SNR, the outage probability can be approximated as follows:
\begin{align}&
\mathrm{P}^o_{1,k} = \sum^{k-1}_{p=0}{k-1 \choose p} (-1)^p K!
\\\nonumber
&   \times\frac{\left[ e^{-\epsilon_{1,k}^*}   \sum_{q=N-M+1}^{\infty}
\frac{(N-M)!\left(\epsilon_{1,k}^*\right)^q }{q!  }
  \right]^{K-k+p+1} }{(K-k)!(k-1)!(K-k+p+1) ((N-M)!)^{K-k+p+1}}
  \\ \nonumber &\approx  \sum^{k-1}_{p=0}{k-1 \choose p} (-1)^p K!
\\\nonumber
&   \times\frac{\left[
\frac{(N-M)!\left(\epsilon_{1,k}^*\right)^{N-M+1}}{(N-M+1)!   }
  \right]^{K-k+p+1} }{(K-k)!(k-1)!(K-k+p+1) ((N-M)!)^{K-k+p+1}}
    \\ \nonumber &\approx  \frac{ K!\left[
\frac{ \left(\epsilon_{1,k}^*\right)^{N-M+1}}{(N-M+1)!  }
  \right]^{K-k+1} }{(K-k)!(k-1)!(K-k+1)  }.
\end{align}
Therefore the theorem is proved.

\section{Proof  for Lemma \ref{lemma 2}}

The sum-rate gap will be evaluated separately for two cases in the following subsections.
\subsection{Case I with $n=1$ and $k=K$} First recall the following integral from Eq. (3.352.4) in \cite{GRADSHTEYN}:
\[
\int^{\infty}_{0}  \frac{1}{1+x\phi} e^{-lx}dx = -\frac{1}{\phi}e^{\frac{l}{\phi}}  \mathbf{E_i}\left(-\frac{l}{\phi}\right)
\]
By using the above result and also Lemma \ref{lemma1}, $\varphi(1,\phi)$ can be expressed as follows:
\begin{align}\label{x1}
\varphi(1,\phi) =&\frac{\phi}{\ln 2}\int^{\infty}_{0}  \frac{1}{1+x\phi}\left(1-\gamma_1 \frac{\left[\gamma(1,x) \right]^{K}}{K}\right)   dx\\ \nonumber= &\frac{1}{\ln 2}   \left(     \sum^{K}_{l=1}{K \choose l}(-1)^l e^{\frac{l}{\phi}}  \mathbf{E_i}\left(-\frac{l}{\phi}\right)  \right)  .
\end{align}

For the worst user, a direct use of Lemma \ref{lemma1} results in a quite complicated expression for $\varphi(K,\phi)$. Instead, we can find a simpler alternative way to calculate this factor, as shown in the following:
\begin{align}\label{x2}
\varphi(K,\phi) =& \int^{\infty}_{0}\log \left(1+x\phi \right) f_{x_K}(x)dx \\ \nonumber =& K \log(e) \int^{\infty}_{0}\ln \left(1+x\phi \right) e^{-Kx}dx
\\ \nonumber =& -  \log(e) e^{\frac{K}{\phi}}\mathbf{E_i}\left(-\frac{K}{\phi}\right).
\end{align}

By substituting \eqref{x1} and \eqref{x2} into the expression for the rate gap, the expression in \eqref{lemma1 eq1} can be obtained.

To obtain the high SNR approximation, first recall that the exponential integral function  has the following series representation \cite{GRADSHTEYN}:
\[
\mathbf{E_i}(x) = \mathbf{C} + \ln(-x) +\sum^{\infty}_{j=1}\frac{x^j}{j\cdots j!},
\] for $x<0$.  Therefore at high SNR,  we have the following approximation:
\begin{align}
\mathcal{E}\left\{\Delta\right\}& \approx -  \log(e)  \left(\mathbf{C}+\ln\left(\frac{K}{\rho}\right)\right) \\ \nonumber &+\frac{2}{ \ln 2}   \left(     \sum^{K}_{l=1}{K \choose l}(-1)^l    \left(\mathbf{C}+\ln\left(\frac{l}{\rho  \alpha_{1,1}^2}\right)\right)  \right)   \\ \nonumber &+2  \log(e) \left(\mathbf{C}+\ln\left(\frac{K}{\rho\alpha_{1,1}^2}\right) \right)  \\ \nonumber &-\frac{1}{ \ln 2}   \left(     \sum^{K}_{l=1}{K \choose l}(-1)^l \left(\mathbf{C}+\ln\left(\frac{l}{\rho}\right)\right)  \right) .
\end{align}

After some manipulations the average gap is given by
 \begin{align}
\frac{\mathcal{E}\left\{\Delta\right\}}{\log e}& \approx     \mathbf{C}+\ln\left(\frac{K}{\rho }\right) -2\ln \alpha^2_{1,1} \\ \nonumber &+         \sum^{K}_{l=1}{K \choose l}(-1)^l    \left(\mathbf{C}+\ln\left(\frac{1}{\rho  }\right)\right)  \\ \nonumber &+         \sum^{K}_{l=1}{K \choose l}(-1)^l    \ln l    +2        \sum^{K}_{l=1}{K \choose l}(-1)^l    \left(-\ln   \alpha_{1,1}^2\right)\\ \nonumber
& =     \mathbf{C}+\ln\left(\frac{K}{\rho }\right) -2\ln \alpha^2_{1,1} +         \sum^{K}_{l=1}{K \choose l}(-1)^l    \ln l   \\ \nonumber &-  \left(\mathbf{C}+\ln\left(\frac{1}{\rho  }\right)\right)+         \sum^{K}_{l=0}{K \choose l}(-1)^l    \left(\mathbf{C}+\ln\left(\frac{1}{\rho  }\right)\right)   \\ \nonumber &-2\left(-\ln   \alpha_{1,1}^2\right)+2        \sum^{K}_{l=0}{K \choose l}(-1)^l    \left(-\ln   \alpha_{1,1}^2\right)
   .
\end{align}
After removing some common factors, the average gap can be simplified  as follows:
 \begin{align}
\frac{\mathcal{E}\left\{\Delta\right\}}{\log e}
& \approx     \mathbf{C}+\ln\left(\frac{K}{\rho }\right) -2\ln \alpha^2_{1,1}  - \left(\mathbf{C}+\ln\left(\frac{1}{\rho  }\right)\right) \\ \nonumber &+         \sum^{K}_{l=1}{K \choose l}(-1)^l    \ln l    +2 \ln   \alpha_{1,1}^2
\\ \nonumber &=  \ln K  +         \sum^{K}_{l=1}{K \choose l}(-1)^l    \ln l
   .
\end{align}
And the first part of the lemma is proved.

\subsection{Case II  with $n=1$ and $k=2$}
It is more complicated to evaluate the average gap for Case II due to the complicated expression for $x_2$. In particular, the factor $\varphi(2,\phi)$ can be expressed as follows:
\begin{align}
\varphi(2,\phi) &=\frac{\phi}{\ln 2}\int^{\infty}_{0}  \frac{1}{1+x\phi}\left(1-\gamma_2     \frac{\left[1-e^{-x} \right]^{K-1}}{K-1}\right. \\ \nonumber &\left. +\gamma_2      \frac{\left[1-e^{-x} \right]^{K}}{K}\right)   dx
\\ \nonumber &=\frac{\phi}{\ln 2}\int^{\infty}_{0}  \frac{1}{1+x\phi}\left(1-K  \sum^{K-1}_{p=0} {K-1 \choose p}(-1)^p e^{-px}  \right. \\ \nonumber &\left. + (K-1)    \sum^{K}_{l=0}{K \choose l}(-1)^l e^{-lx}  \right)   dx.
\end{align}
Since $\int^{\infty}_{0}  \frac{1}{1+x\phi}dx \rightarrow \infty$, therefore it is important to  remove the factors in the integral related to $\frac{1}{1+x\phi}$ in order to facilitate the high SNR approximation. Motivated by this,   the factor $\varphi(2,\phi)$ can be rewritten  as follows:
\begin{align}\nonumber
\varphi(2,\phi)=&\frac{\phi}{\ln 2}\int^{\infty}_{0}  \frac{1}{1+x\phi}\left(-K  \sum^{K-1}_{p=1} {K-1 \choose p}(-1)^p e^{-px}  \right. \\   &\left. + (K-1)    \sum^{K}_{l=1}{K \choose l}(-1)^l e^{-lx}  \right)   dx.
\end{align}
Following the   steps as those used  in the previous section, the integral in the above equation can be evaluated and we can have the following:
\begin{align}\label{x3}
\varphi(2,\phi) &=\frac{1}{\ln 2}   \left(K  \sum^{K-1}_{p=1} {K-1 \choose p}(-1)^p   e^{\frac{p}{\phi}}  \mathbf{E_i}\left(-\frac{p}{\phi}\right) \right. \\ \nonumber &\left. - (K-1)    \sum^{K}_{l=1}{K \choose l}(-1)^l e^{\frac{l}{\phi}}  \mathbf{E_i}\left(-\frac{l}{\phi}\right)  \right)  .
\end{align}
Substituting \eqref{x3} and \eqref{x1} into \eqref{lemma eq}, the exact exact expression of the average rate gap can be obtained as in the lemma.

At high SNR, the exponential integral function can be simplified as discussed previously, and   the average rate gap can be approximated as follows:
\begin{align}\nonumber
\mathcal{E}\left\{\Delta\right\}  &\approx\frac{1}{ \ln 2}   \left(K  \sum^{K-1}_{p=1} {K-1 \choose p}(-1)^p     \left(\mathbf{C}+\ln\left(\frac{p}{\rho}\right)\right)\right. \\ \nonumber &\left.  - (K-1)    \sum^{K}_{l=1}{K \choose l}(-1)^l \left(\mathbf{C}+\ln\left(\frac{l}{\rho}\right)\right)  \right)   dx \\ \nonumber &+\frac{2}{ \ln 2}   \left(     \sum^{K}_{l=1}{K \choose l}(-1)^l  \left(\mathbf{C}+\ln\left(-\frac{l}{\rho  \alpha_{1,1}^2}\right) \right) \right)   \\ \nonumber &-  \frac{2}{\ln 2}   \left(K  \sum^{K-1}_{p=1} {K-1 \choose p}(-1)^p    \left(\mathbf{C}+\ln\left(-\frac{p}{\rho  \alpha_{1,1}^2}\right) \right) \right.\\ \nonumber &\left.- (K-1)    \sum^{K}_{l=1}{K \choose l}(-1)^l  \left(\mathbf{C}+\ln\left(-\frac{l}{\rho  \alpha_{1,1}^2}\right)\right)  \right)    \\   &-\frac{1}{ \ln 2}   \left(     \sum^{K}_{l=1}{K \choose l}(-1)^l  \left(\mathbf{C}+\ln\left(-\frac{l}{\rho}\right) \right) \right) .
\end{align}
With some algebraic manipulations, the average rate gap can be expressed as \eqref{fig2eq}.
\begin{figure*}
\begin{align}\label{fig2eq}
\mathcal{E}\left\{\Delta\right\}  &\approx \frac{1}{ \ln 2}   \left(K  \sum^{K-1}_{p=1} {K-1 \choose p}(-1)^p      \ln p  - (K-1)    \sum^{K}_{l=1}{K \choose l}(-1)^l \ln l  \right) +\frac{2}{ \ln 2}   \left(     \sum^{K}_{l=1}{K \choose l}(-1)^l   \ln l \right)   \\ \nonumber &-  \frac{2}{ \ln 2}   \left(K  \sum^{K-1}_{p=1} {K-1 \choose p}(-1)^p   \ln p - (K-1)    \sum^{K}_{l=1}{K \choose l}(-1)^l  \ln l  \right)    +\frac{1}{ \ln 2}   \left(-     \sum^{K}_{l=1}{K \choose l}(-1)^l   \ln l \right) .
\end{align}
\end{figure*}
After   those common factors in \eqref{fig2eq} are removed, the high SNR approximation shown in \eqref{hign snr part 2} can be obtained, and the lemma is proved.

\section{Proof for Lemma \ref{lemma 3}}
Without loss of generality,  we will take the users in the first cluster as an example.
Recall that the $n$-th user, $n<k$,  in the first cluster   tries to decode the $k$-th user's message with the following SINR: $$SINR^k_{1,n}=\frac{|\mathbf{v}_{1,n}^H\mathbf{h}_{1,1n}|^2  \alpha_{1,k}^2 }{ |\mathbf{v}_{1,n}^H\mathbf{h}_{1,1n} |^2 \alpha_{1,n}^2 + \frac{1}{\rho}}.$$ If successful, the $n$-th user will decode its own message with the following SINR:
\begin{align}
SINR^n_{1,n} &=   \rho |\mathbf{v}_{1,n}^H\mathbf{H}_{1,1}\mathbf{p}_1 |^2 \alpha_{1,n}^2
\\\nonumber &=   \rho |\mathbf{v}_{1,n}^H\mathbf{H}_{1,1}\mathbf{p}_1 |^2 \left(1-  \frac{ \epsilon_{1,k}\left( |\mathbf{v}_{1,k}^H\mathbf{h}_{1,1k}|^2 +\frac{1}{\rho}\right)}{|\mathbf{v}_{1,k}^H\mathbf{h}_{1,1k}|^2 (1+\epsilon_{1,k})}\right),
\end{align}
if $\alpha_{1,n}^2 >0$.
Again with the same notation used in the proof of Theorem \ref{theorem1}, the SINR at the $n$-th user can be expressed as follows:
\begin{align}\label{sinr1}
SINR^n_{1,n} &=     \rho x_n \left(1-  \frac{ \epsilon_{1,k}\left( x_k +\frac{1}{\rho}\right)}{x_k (1+\epsilon_{1,k})}\right),
\end{align}
if $\alpha_{1,n}^2 >0$.

Thus there are two conditions before the SINR expression in \eqref{sinr1} can be used. One is $\alpha_{1,n}^2 >0$ and the other is that the $n$-th user can decode the $k$-th user's message, i.e., $\log(1+SINR^k_{1,n})>R_{1,k}$.
Therefore the outage events at the $n$-th user can be categorized into three following types:
\begin{enumerate}
\item Events with $\alpha_{1,n}^2 =0$, which means $1\leq  \frac{ \epsilon_{1,k}\left( x_k +\frac{1}{\rho}\right)}{x_k (1+\epsilon_{1,k})}$ or equivalently
\begin{align}
x_k\leq \frac{\epsilon_{1,k}}{\rho}.
\end{align}

\item Events with $\alpha_{1,n}^2 >0$ and $\log(1+SINR^k_{1,n})<R_{1,k}$.

\item Events with $\alpha_{1,n}^2 >0$, $\log(1+SINR^k_{1,n})>R_{1,k}$ and $SINR^n_{1,n}<\epsilon_{1,n}$.
\end{enumerate}

Because $|\mathbf{v}_{1,k}^H\mathbf{h}_{1,1k}|^2<|\mathbf{v}_{1,n}^H\mathbf{h}_{1,1n}|^2$, it is straightforward to show $SINR^k_{1,n}>SINR^k_{1,k}$, which means
\begin{align}
&\mathrm{P}\left(\alpha_{1,n}^2 >0,\log(1+SINR^k_{1,n})<R_{1,k}\right)\\ \nonumber &=\mathcal{E}_{0< \alpha_{1,n}^2\leq 1}\left\{\mathrm{P}\left( SINR^k_{1,n}<\epsilon_{1,k}\right)\right\})\\ \nonumber &=\mathcal{E}_{0< \alpha_{1,n}^2\leq 1}\left\{\mathrm{P}\left(SINR^k_{1,k}=\epsilon_{1,k},SINR^k_{1,n}<\epsilon_{1,k}\right)\right\}=0,
\end{align}
i.e., the $n$-th user can decode the $k$-th user's information as long as the $k$-th user can decode its own. But if $\alpha_{1,k}^2 =1$, i.e., the BS allocates all the power to the $k$-th  user, outage will occur at the $n$-th user.

Therefore the outage probability experienced by the $n$-th user is given by
\begin{align}\nonumber
\mathrm{P}^o_n &= \mathrm{P}\left(    \rho x_n \left(1-  \frac{ \epsilon_{1,k}\left( x_k +\frac{1}{\rho}\right)}{x_k (1+\epsilon_{1,k})}\right)<\epsilon_{1,n}, x_k> \frac{\epsilon_{1,k}}{\rho}\right) \\ &+\mathrm{P}\left(x_k\leq \frac{\epsilon_{1,k}}{\rho}\right),\label{outage x1}
\end{align}
which follows from the following simplification:
\begin{align}
&\nonumber
\mathrm{P}\left(\alpha_{1,n}^2 >0,\log(1+SINR^k_{1,n})>R_{1,k},SINR^n_{1,n}<\epsilon_{1,n}\right)
\\   &=
\mathrm{P}\left(\alpha_{1,n}^2 >0, SINR^n_{1,n}<\epsilon_{1,n}\right).
\end{align}

Define the first factor in the expression for the outage probability in \eqref{outage x1} by $Q_2\triangleq  \mathrm{P}\left(    \rho x_n \left(1-  \frac{ \epsilon_{1,k}\left( x_k +\frac{1}{\rho}\right)}{x_k (1+\epsilon_{1,k})}\right)<\epsilon_{1,n}, x_k> \frac{\epsilon_{1,k}}{\rho}\right)$. This factor can be evaluated as follows:
\begin{align}
Q_2 &= \mathrm{P}\left(      x_n \left(   \frac{ x_k- \frac{\epsilon_{1,k}}{\rho} }{x_k }\right)<\tilde{\epsilon}_{1,n}, x_k> \frac{\epsilon_{1,k}}{\rho}\right) \\  &= \mathcal{E}_{x_k}\left\{\mathrm{P}\left(    x_k<  x_n <\frac{\tilde{\epsilon}_{1,n}x_k}{ x_k- \frac{\epsilon_{1,k}}{\rho} }\right)\right\}\label{exp}.
\end{align}
where $\tilde{\epsilon}_{1,n}=\frac{(1+\epsilon_{1,k})\epsilon_{1,n}}{\rho}$. It is important to note that the expectation  in \eqref{exp} is taken over the following range 
\[
 \frac{\epsilon_{1,k}}{\rho}<x_k< \frac{\epsilon_{1,k}}{\rho} +\tilde{\epsilon}_{1,n},
\]
where the upper bound is due to the constraint $ x_k<   \frac{\tilde{\epsilon}_{1,n}x_k}{ x_k- \frac{\epsilon_{1,k}}{\rho} }$.

Now the factor $Q_2$ can be evaluated as follows:
\begin{align}
Q_2   &= \mathcal{E}_{x_k}\left\{    F_{x_n}\left(\frac{\tilde{\epsilon}_{1,n}x_k}{ x_k- \frac{\epsilon_{1,k}}{\rho} }  \right) -F_{x_n}(x_k) \right\}\\ \nonumber & =  \int^{\frac{\epsilon_{1,k}}{\rho} +\tilde{\epsilon}_{1,n}}_{ \frac{\epsilon_{1,k}}{\rho}}    \left( F_{x_n}\left(\frac{\tilde{\epsilon}_{1,n} y}{ y- \frac{\epsilon_{1,k}}{\rho} }  \right) -F_{x_n}(y)   \right)f_{x_k}(y)dy.
\end{align}
At high SNR,  $Q_2$ can be upper bounded as follow:
\begin{align}
Q_2   &<  \int^{\frac{\epsilon_{1,k}}{\rho} +\tilde{\epsilon}_{1,n}}_{ \frac{\epsilon_{1,k}}{\rho}}   f_{{x}_k}(y) dy,
\end{align}
since $F_{x_n}\left(\frac{\tilde{\epsilon}_{1,n} y}{ y- \frac{\epsilon_{1,k}}{\rho} }  \right) -F_{x_n}(y)  \leq F_{x_n}\left(\frac{\tilde{\epsilon}_{1,n} y}{ y- \frac{\epsilon_{1,k}}{\rho} }  \right) \leq 1$.

With this upper bound, the overall outage probability can be upper bounded as follows:
\begin{align}\label{outage x2}
\mathrm{P}^o_n &=Q_2+\mathrm{P}\left(x_k\leq \frac{\epsilon_{1,k}}{\rho}\right)\\ \nonumber &\leq\int^{\frac{\epsilon_{1,k}}{\rho} +\tilde{\epsilon}_{1,n}}_{ \frac{\epsilon_{1,k}}{\rho}}   f_{{x}_k}(y) dy+\mathrm{P}\left(x_k\leq \frac{\epsilon_{1,k}}{\rho}\right)\\ \nonumber &= F_{{x}_n}\left(\frac{\epsilon_{1,k}}{\rho} +\tilde{\epsilon}_{1,n}\right).
\end{align}
To find a high SNR approximation of $F_{{x}_n}(x)$, first recall that  the CDF of $x_n$ is given by
\begin{eqnarray}
F_{{x}_n}(x) = \gamma_n \sum^{n-1}_{j=0}{n-1 \choose j} (-1)^j \frac{ [F_{\tilde{x}_n}(x) ]^{K-n+j+1} }{K-n+j+1}.
\end{eqnarray}
When $x\rightarrow 0$, we have
\begin{eqnarray}
 F_{\tilde{x}_n}(x)  = \frac{\gamma(N-M+1,x)}{(N-M)!} \approx \frac{x^{N-M+1}}{(N-M+1)!}.
\end{eqnarray}

Therefore, the CDF of $x_n$ can be approximated as follows:
\begin{align}
F_{{x}_n}(x) &\nonumber\approx \gamma_n \sum^{n-1}_{j=0}{n-1 \choose j} (-1)^j \frac{ \left(\frac{x^{N-M+1}}{(N-M+1)!}\right)^{K-k+j+1} }{K-n+j+1}\\  &
\approx   \frac{ \gamma_n}{K-n+1}\left(\frac{x^{N-M+1}}{(N-M+1)!}\right)^{K-k+1} ,
\end{align}
when $x\rightarrow 0$.

Substituting the above approximation into \eqref{outage x2},  the overall outage probability  can be upper bounded as follows:
\begin{align}
\mathrm{P}^o_n &\leq   \frac{ \gamma_n}{K-n+1}\left(\frac{\left(\frac{\epsilon_{1,k}}{\rho} +\tilde{\epsilon}_{1,n}\right)^{N-M+1}}{(N-M+1)!}\right)^{K-k+1} \\ \nonumber &\rightarrow \frac{1}{\rho^{(N-M+1)(K-k+1)}}.
\end{align}
And the proof is completed.

\section{Proof for Lemma \ref{theorem2}}
With the same notations used in the proof for Theorem \ref{theorem1}, the SINR at the $n$-th user can be expressed as follows:
\begin{align}
SINR^n_{1,n} &=     \rho x_n \frac{\sqrt{1+\rho x_k} -1}{\rho x_k},
\end{align}
if it can decode the $k$-th user's message, i.e.,
\begin{align}
\log\left(1+\frac{x_n \alpha_{1,k}^2 }{ x_n \alpha_{1,n}^2 + \frac{1}{\rho}}\right) >R_{1,k}.
\end{align}
  Therefore the outage events at the $n$-th user can be categorized into the two following types:
\begin{itemize}
\item Events in which  the $n$-th user cannot decode the $k$-th user, i.e.,  $$\log\left(1+\frac{x_n \alpha_{1,k}^2 }{ x_n \alpha_{1,n}^2 + \frac{1}{\rho}}\right) <R_{1,k} $$
\item Events  in which the $n$-th user can decode the $k$-th user, but cannot decode its own, i.e.,
$$\log\left(1+\frac{x_n \alpha_{1,k}^2 }{ x_n \alpha_{1,n}^2 + \frac{1}{\rho}}\right) >R_{1,k},$$ and $$ \log\left(1+\rho x_n\alpha_{1,n}^2<R_{1,n}\right) .$$

\end{itemize}

Therefore the outage probability experienced by the $n$-th user is given by
{\small \begin{align}\label{eqx6}
\mathrm{P}^o_n &= \mathrm{P}\left( \log\left( \frac{x_n\rho+1}{ x_n \alpha_{1,n}^2\rho + 1}\right) <R_{1,k}  \right)\\ \nonumber &+ \underset{Q_3}{\underbrace{\mathrm{P}\left( \log\left( \frac{x_n\rho+1}{ x_n \alpha_{1,n}^2\rho + 1}\right) >R_{1,k},\log\left(1+\rho x_n\alpha_{1,n}^2<R_{1,n}\right) \right)}}.
\end{align}}
The first factor of $\mathrm{P}^o_n $ can be calculated as follows:
 \begin{align}
 &Q_4\triangleq  \mathrm{P}\left( \log\left( \frac{x_n\rho+1}{ x_n \alpha_{1,n}^2\rho + 1}\right) <R_{1,k}  \right) \\ \nonumber& =\left\{\begin{array}{ll} \mathrm{P}\left( x_k<x_n<\frac{2^{R_{1,k}}-1}{\rho(1-2^{R_{1,k}}\alpha_{1,n})}  \right), & \text{if}  \quad 1>2^{R_{1,k}}\alpha_{1,n}\\ 1, & \text{otherwise} \end{array}\right..
\end{align}

The constraint of $1>2^{R_{1,k}}\alpha_{1,n}$ is equivalent to the following one:
\begin{align}
x_k>\frac{2^{R_{1,k}}(2^{R_{1,k}}-2)}{\rho}.
\end{align}
As a result, $\mathrm{P}\left( \log\left( \frac{x_n\rho+1}{ x_n \alpha_{1,n}^2\rho + 1}\right) <R_{1,k}  \right)$ can be expressed as follows:{\small
 \begin{align}
  Q_4 = \left\{\begin{array}{lc} \mathrm{P}\left( x_k<x_n<\frac{2^{R_{1,k}}-1}{\rho(1-2^{R_{1,k}}\alpha_{1,n})}  \right., & \text{if}  \quad 1>2^{R_{1,k}}\alpha_{1,n}\\ \left.x_k>\frac{2^{R_{1,k}}(2^{R_{1,k}}-2)}{\rho}\right) & \& \quad R_{1,k}>1\\
 \mathrm{P}\left( x_k<x_n<\frac{2^{R_{1,k}}-1}{\rho(1-2^{R_{1,k}}\alpha_{1,n})}  \right), & \text{if}  \quad 1>2^{R_{1,k}}\alpha_{1,n}\\   & \& \quad R_{1,k}\leq 1 \\ 1, & \text{otherwise} \end{array}\right..
\end{align}}

Consequently $\mathrm{P}\left( \log\left( \frac{x_n\rho+1}{ x_n \alpha_{1,n}^2\rho + 1}\right) <R_{1,k}  \right)$ can be upper bounded as follows:
 \begin{align}\label{q42}
 Q_4& \leq  \mathrm{P}\left( x_k<x_n<\frac{2^{R_{1,k}}-1}{\rho(1-2^{R_{1,k}}\alpha_{1,n})}  \right) \\ \nonumber &+ \mathrm{P}\left(   x_k<\frac{2^{R_{1,k}}(2^{R_{1,k}}-2)}{\rho} \right),
\end{align}
if $R_{1,k}>1$, otherwise
 \begin{align}\label{q41}
Q_4& =  \mathrm{P}\left( x_k<x_n<\frac{2^{R_{1,k}}-1}{\rho(1-2^{R_{1,k}}\alpha_{1,n})}  \right) .
\end{align}
Comparing  \eqref{q42} to \eqref{q41}, one can observe that the probability   $\mathrm{P}\left(   x_k<\frac{2^{R_{1,k}}(2^{R_{1,k}}-2)}{\rho} \right)$ does not need to be taken into consideration for the case of $0\leq R_{1,k}\leq 1$. In the following, we first focus on the case   $R_{1,k}>1$.

It is important to note that the constraint $x_k<x_n<\frac{2^{R_{1,k}}-1}{\rho(1-2^{R_{1,k}}\alpha_{1,n})} $ yields the following additional constraint for $x_k$:
\begin{align}
x_k< \frac{2^{R_{1,k}}-1}{\rho\left(1-2^{R_{1,k}}\frac{\sqrt{1+\rho x_k} -1}{\rho x_k}\right)} ,
\end{align}
which leads to the following inequality:
\begin{align}
x_k< \frac{2^{2R_{1,k}}-1}{\rho}.
\end{align}

Therefore $\mathrm{P}\left( \log\left( \frac{x_n\rho+1}{ x_n \alpha_{1,n}^2\rho + 1}\right) <R_{1,k}  \right)$ can be upper bounded as follows:
 \begin{align}\label{eqx1}
 & \mathrm{P}\left( \log\left( \frac{x_n\rho+1}{ x_n \alpha_{1,n}^2\rho + 1}\right) <R_{1,k}  \right) \\ \nonumber& \leq  \mathrm{P}\left( x_k<x_n<\frac{2^{R_{1,k}}-1}{\rho(1-2^{R_{1,k}}\alpha_{1,n})}  ,x_k< \frac{2^{2R_{1,k}}-1}{\rho}\right) \\ \nonumber &+ \mathrm{P}\left(   x_k<\frac{2^{R_{1,k}}(2^{R_{1,k}}-2)}{\rho} \right)\\ \nonumber& \leq  \mathrm{P}\left(x_k< \frac{2^{2R_{1,k}}-1}{\rho}\right)  + \mathrm{P}\left(   x_k<\frac{2^{R_{1,k}}(2^{R_{1,k}}-2)}{\rho} \right).
\end{align}

Following     steps similar to those used in the previous section, we   have the following asymptotic result:
 \begin{align}  \mathrm{P}\left( x_k< \frac{2^{2R_{1,k}}-1}{\rho} \right) \rightarrow  \frac{1}{\rho^{(N-M+1)(K-k+1)}}.\label{eqx3}
\end{align}
The other probabilities have the same asymptotic behavior, and therefore by combining \eqref{eqx1}  and \eqref{eqx3}, we have
{\small \begin{align}\label{eqx4}
  \mathrm{P}\left( \log\left( \frac{x_n\rho+1}{ x_n \alpha_{1,n}^2\rho + 1}\right) <R_{1,k}  \right)  \dot\leq  \frac{1}{\rho^{(N-M+1)(K-k+1)}},
\end{align} }
where  $a \dot\leq b$ denotes $\underset{\rho\rightarrow \infty}{\lim}\frac{\log a}{\log \rho}\geq \underset{\rho\rightarrow \infty}{\lim}\frac{\log b}{\log \rho}$ \cite{Zhengl03}. The above conclusion is also valid for the case   $0\leq R_{1,k}\leq 1$.

The factor $Q_3$ can be upper bounded as follows:
\begin{align}
Q_3 &\leq     \mathrm{P}\left(  \log\left(1+\rho x_n\alpha_{1,n}^2<R_{1,n}\right)\right)\\ \nonumber &=     \mathrm{P}\left(  x_k< x_n<\frac{2^{R_{1,n}}-1}{\rho \alpha_{1,n}^2} \right)
\end{align}
A hidden constraint on $x_k$ due to $ x_k< x_n<\frac{2^{R_{1,n}}-1}{\rho \alpha_{1,n}^2}$ is
 \begin{align}
 x_k< \frac{2^{R_{1,n}}-1}{\rho \frac{\sqrt{1+\rho x_k} -1}{\rho x_k}},
\end{align}
which yields $x_k<\frac{2^{2R_{1,n}}-1}{\rho}$.

Therefore the factor $Q_3$ can be further upper bounded as follows:
\begin{align}\label{eqx5}
Q_3 &\leq        \mathrm{P}\left(  x_k< x_n<\frac{2^{R_{1,n}}-1}{\rho \alpha_{1,n}^2}, x_k<\frac{2^{2R_{1,n}}-1}{\rho} \right)\\ \nonumber
&\leq        \mathrm{P}\left(   x_k<\frac{2^{2R_{1,n}}-1}{\rho} \right)\rightarrow \frac{1}{\rho^{(N-M+1)(K-k+1)}}.
\end{align}

Combing  \eqref{eqx4},   \eqref{eqx5} and \eqref{eqx6},  the overall outage probability can be upper bounded as follows:
\begin{align}\label{eqx6}
\mathrm{P}^o_n  \rightarrow \frac{1}{\rho^{(N-M+1)(K-k+1)}}
\end{align}
And the proof is completed.
 \bibliographystyle{IEEEtran}
\bibliography{IEEEfull,trasfer}
  \end{document}